\documentclass[showpacs,preprintnumbers,amsmath,amssymb,twocolumn]{revtex4}

\usepackage{graphicx}
\usepackage{dcolumn}
\usepackage{txfonts,bm}
\usepackage{hyperref}
\usepackage{color}
\usepackage{epsfig}
\usepackage{multirow}
\usepackage{subfigure}
\usepackage{amssymb}
\usepackage{amsmath}

\begin{document}

\title{The recoil correction and spin-orbit force for the possible
$B^* \bar{B}^{*}$ and $D^* \bar{D}^{*}$ states}

\author{Lu Zhao$^{1}${\footnote{Email: Luzhao@pku.edu.cn}},
Li Ma$^{1}${\footnote{Email: lima@pku.edu.cn}}, Shi-Lin
Zhu$^{1,2}${\footnote{Email: zhusl@pku.edu.cn}}}

\affiliation{$^{1}$ Department of Physics and State Key Laboratory
of Nuclear Physics and Technology, Peking University, Beijing
100871, China\\
$^2$Collaborative Innovation Center of Quantum Matter, Beijing
100871, China }

\begin{abstract}
In the framework of the one-boson exchange model, we have calculated
the effective potentials between two heavy mesons $B^* \bar{B}^{*}$
and $D^* \bar{D}^{*}$ from the t- and u-channel $\pi$-, $\eta$-,
$\rho$-, $\omega$- and $\sigma$-meson exchanges. We keep the recoil
corrections to the $B^* \bar{B}^{*}$ and $D^* \bar{D}^{*}$ systems
up to $O(\frac{1}{M^2})$, which turns out to be important for the
very loosely bound molecular states. Our numerical results show that
the momentum-related corrections are favorable to the formation of
the molecular states in the $I^G=1^+$, $J^{PC}=1^{+-}$ in the $B^*
\bar{B}^{*}$ and $D^* \bar{D}^{*}$ systems.

\end{abstract}

 \maketitle{}

\section{INTRODUCTION}\label{introduction}

A lot of charmonium-like states have been reported in the past
decade by the experiment collaborations such as Belle, $Barbar$,
CDF, D0, LHCb, BESIII, and CLEOc. The underlying structures of many
charmonium-like states are not very clear. Sometimes they are called
as XYZ states. They decay into conventional chromium, but not all of
them can be accommodated into the quark-model charmonium spectrum.
The neutral XYZ states include $X(3872)$ \cite{Choi:2003}, $Y(4260)$
\cite{B. Aubert:2005}, $Y(4008)$ \cite{C.Z. Yuan:2007}, $Y(4360)$
\cite{B. Aubert:2007}, $Y(4660)$ \cite{X.L. Wang:2007}, and
$Y(4630)$ \cite{G.Pakhlova:2008} etc. There are also many charged
charmonium-like states such as $Z_1(4050)$ and $Z_2(4250)$
\cite{R.Mizuk:2008}, $Z_c(4485)$ \cite{Choi:2008, K.Chilikin:2013},
$Z_c(3900)$ \cite{Ablikim:2013, Z.Q. Liu:2013-2, T.Xiao:2013},
$Z_c(4020)$ \cite{M. Ablikim:2013}, $Z_c(4025)$ \cite{Ablikim:2014}.
The charged bottomonium-like states $Z_b(10610)$ and $Z_b(10650)$
were observed by Belle Collaboration \cite{I.Adachi:2011}.

Theoretical speculations of these XYZ states include the hybrid
meson \cite{ZHUPLB2005}, tetraquark states \cite{H.Hogaasen:2006,
D.Ebert:2006, N.Barnea:2006, Y.Cui:2007, R.D.Matheus:2007,
T.W.Chiu:2007, L.Zhao:2014}, dynamically generated resonance
\cite{D.Gamermann:2007} and molecular states \cite{F.E.Close:2004,
M.B.Voloshin:2004, C.Y.Wong:2004, E.S.Swanson:2004,
N.A.Tornqvist:2004, Y.R.Liu:2010, N.Li:2012, mali, liuxiaohai} etc.
Since many of these XYZ states are close to the thresholds of a pair
of charmed or bottom mesons, the molecular hypothesis seems a
natural picture for some of these states.

Within the framework of the molecular states, there exist extensive
investigations of the charged $Z_c$ and $Z_b$ states
\cite{Q.Wang:2013, Oset, Y.R.Liu:2008, X.Liu:2009, A.E.Bondar:2011,
L.Xiang:2011, W.Chen:2014, J.He:2013, Z.F.Sun:2012, C.D. Deng:2014,
J.M. Dias:2014}. In our previous work \cite{L.Zhao:2013}, we explored
the possibility of $Z_c(3900)$ as the isovector molecule partner of
$X(3872)$ and considered the recoil correction and the spin-orbit
force in the $D \bar{D}^*$ and $B \bar{B}^*$ system very carefully.

Although there exist quite a few literatures on the possibility of
$Z_c(4025)$ as the $D^* \bar{D}^*$ molecular state and $Z_b(10650)$
as the $B^* \bar{B}^*$ molecular state, most of the available
investigations are either based on the heavy quark spin-flavor
symmetry or derived in the $m_Q\to \infty$. In other words, the
recoil correction and the spin-orbit interaction have not been
investigated for the $D^* \bar{D}^*$ and $B^* \bar{B}^*$ systems.
Since the binding energies of these system are very small, the high
order recoil correction and the spin-orbit interaction may lead to
significant corrections.

In this work, we will go one step further. We will consider the
recoil correction and the spin-orbit force for the $D^* \bar{D}^*$
and $B^* \bar{B}^*$ systems. With the one-boson-exchange model
(OBE), we will derive the effective potential with the relativistic
Lagrangian and keep the momentum related terms explicitly in order
to derive the recoil correction and the spin-orbit interaction up to
$O(1/M^2)$, where $M$ is the mass of the heavy meson. We investigate
the $B^* \bar{B}^*$ system with $I^G=1^+, J^P=1^{+-}$ for
$Z_b(10650)$, and the $D^* \bar{D}^*$ system with $I^G=1^+,
J^P=1^{+-}$ for $Z_c(4025)$. For completeness, we also investigate
the $B^* \bar{B}^*$ system and $D^* \bar{D}^*$ system with other
quantum numbers: $I^G=1^-, J^P=1^{++}$, $I^G=0^+, J^P=1^{++}$, and
$I^G=0^-, J^P=1^{+-}$.  Compared to the $D \bar{D}^*$ case, the
expressions of the recoil corrections and spin orbit force are more
complicated. There appear several new structures. For some systems,
the numerical results show that the high order correction is
important for the loosely bound heavy-meson states.

This paper is organized as follows. We first introduce the formalism
of the derivation of the effective potential in Sec.
\ref{potential}. We present our numerical results in Sec.
\ref{Numerical-BB} and Sec. \ref{Numerical-DD}. The last section is
the summary and discussion.

\section{The effective potential}\label{potential}
\begin{center}
\textbf{A. Wave function, Effective Lagrangian and Coupling
constants}
\end{center}

First, we construct the flavor wave functions of the isovector and
isoscalar molecular states composed of the $B^*\bar{B}^{*}$ and
$D^*\bar{D}^{*}$ as in Refs. \cite{Y.R.Liu:2008,X.Liu:2009}. The
flavor wave function of the $B^*\bar{B}^{*}$ system reads
\begin{equation}
\left\{
  \begin{array}{ll}
    |1,1\rangle = |B^{*+}\bar{B}^{*0}\rangle ,  \\
    |1,-1\rangle = |B^{*-}B^{*0}\rangle , \\
    |1,0\rangle = \frac{1}{\sqrt{2}}(|B^{*+}B^{*-}\rangle
-|B^{*0}\bar{B}^{*0}\rangle) ,
  \end{array}
\right.
\end{equation}

\begin{equation}
|0,0\rangle = \frac{1}{\sqrt{2}}(|B^{*+}B^{*-}\rangle
+|B^{*0}\bar{B}^{*0}\rangle)
\end{equation}
For the $D^*\bar{D}^{*}$ system
\begin{equation}
\left\{
  \begin{array}{ll}
    |1,1\rangle =|\bar{D}^{*0}D^{*+}\rangle ,  \\
    |1,-1\rangle = |D^{*0}D^{*-}\rangle ,  \\
    |1,0\rangle = \frac{1}{\sqrt{2}}(|\bar{D}^{*0}D^{*0}\rangle
-|D^{*-}D^{*+}\rangle) ,
  \end{array}
\right.
\end{equation}

\begin{equation}
|0,0\rangle = \frac{1}{\sqrt{2}}(|\bar{D}^{*0}D^{*0}\rangle
+|D^{*-}D^{*+}\rangle)
\end{equation}

\begin{figure}[ht]
  \begin{center}
    \rotatebox{0}{\includegraphics*[width=0.30\textwidth]{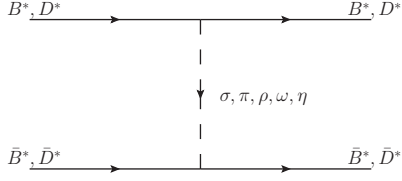}}
    \caption{ The Feynman diagrams for both the $D^*\bar{D}^{*}$ and
$B^*\bar{B}^{*}$ systems at the tree level.} \label{Feynman-digram}
  \end{center}
\end{figure}

The meson exchange Feynman diagrams for the $B^*\bar{B}^{*}$ and
$D^*\bar{D}^{*}$ systems at the tree level is shown in Fig. \ref{Feynman-digram}.

Based on the chiral symmetry, the Lagrangian for the pseudoscalar,
scalar and vector meson interaction with the heavy flavor mesons
reads
\begin{eqnarray}
\mathcal{L}_{P}&=&-i\frac{2g}{f_\pi}\bar{M}
P^{*\mu}_b\partial_{\mu}\phi_{ba}P^{\dag}_{a}+i\frac{2g}{f_\pi}\bar{M} P_b\partial_{\mu}\phi_{ba}P^{*\mu\dag}_{a} \nonumber\\
&-& \frac{g}{f_\pi}
P^{*\mu}_b\partial^{\alpha}\phi_{ba}\partial^{\beta}P^{*\nu\dag}_{a}\epsilon_{\mu\nu\alpha\beta}
+ \frac{g}{f_\pi}
\partial^{\beta}P^{*\mu}_b\partial^{\alpha}\phi_{ba}P^{*\nu\dag}_{a}\epsilon_{\mu\nu\alpha\beta},\label{pseudo-exchange}
\end{eqnarray}

\begin{eqnarray}
\widetilde{\mathcal{L}_{P}}&=&-i\frac{2g}{f_\pi}\bar{M}\widetilde{P^{\dag}_{a}}
\partial_{\mu}\phi_{ab}\widetilde{P^{*\mu}_b}-i\frac{2g}{f_\pi}\bar{M}\widetilde{P^{*\mu\dag}_{a}}\partial_{\mu}\phi_{ab}\widetilde{P_b} \nonumber\\
&+& \frac{g}{f_\pi}
\partial^{\beta}\widetilde{P^{*\mu\dag}_{a}}\partial^{\alpha}\phi_{ab}\widetilde{P^{*\nu}_b}\epsilon_{\mu\nu\alpha\beta}
-\frac{g}{f_\pi}
\widetilde{P^{*\mu\dag}_{a}}\partial^{\alpha}\phi_{ab}\partial^{\beta}\widetilde{P^{*\nu}_b}\epsilon_{\mu\nu\alpha\beta},\label{anti-pseudo-exchange}
\end{eqnarray}

\begin{eqnarray}
\mathcal{L}_{V}&=&i\frac{\beta g_v}{\sqrt{2}} P_b
V^{\mu}_{ba}\partial_{\mu}P^{\dag}_{a}-i\frac{\beta g_v}{\sqrt{2}}
\partial_{\mu}P_b V^{\mu}_{ba}P^{\dag}_{a} \nonumber\\
&-&i\sqrt{2}\lambda g_v\epsilon_{\mu\alpha\beta\nu}\partial^{\mu}P_b
\partial^{\alpha}V^{\beta}_{ba}P^{*\nu\dag}_{a}\nonumber\\
&-&i\sqrt{2}\lambda g_v \epsilon_{\mu\alpha\beta\nu}P^{*\mu}_b
\partial^{\alpha}V^{\beta}_{ba}\partial^{\nu}P^{\dag}_{a}\nonumber\\
&-&i\frac{\beta g_v}{\sqrt{2}} P^{*\nu}_b
V^{\mu}_{ba}\partial_{\mu}P^{*\dag}_{\nu a}+ i\frac{\beta
g_v}{\sqrt{2}} \partial_{\mu}P^{*\nu}_b V^{\mu}_{ba}P^{*\dag}_{\nu a}\nonumber\\
&-&i2\sqrt{2}\lambda g_v \bar{M^*}P^{*\mu}_b
(\partial_{\mu}V_{\nu}-\partial_{\nu}V_{\mu})_{ba}P^{*\nu\dag}_a,\label{vector-exchange}
\end{eqnarray}

\begin{eqnarray}
\widetilde{\mathcal{L}_{V}}&=&-i\frac{\beta g_v}{\sqrt{2}}
\partial_{\mu}\widetilde{P^{\dag}_{a}} V^{\mu}_{ab}\widetilde{P_b}+
i\frac{\beta g_v}{\sqrt{2}} \widetilde{P^{\dag}_{a}} V^{\mu}_{ab}\partial_{\mu}\widetilde{P_b} \nonumber\\
&+&i\sqrt{2}\lambda g_v\epsilon_{\mu\alpha\beta\nu}
\widetilde{P^{*\mu\dag}_{a}}\partial^{\alpha}V^{\beta}_{ab}\partial^{\nu}\widetilde{P_b}\nonumber\\
&+&i\sqrt{2}\lambda g_v \epsilon_{\mu\alpha\beta\nu}\partial^{\mu}
\widetilde{P^{\dag}_{a}}\partial^{\alpha}V^{\beta}_{ab}\widetilde{P^{*\nu}_b}\nonumber\\
&+&i\frac{\beta g_v}{\sqrt{2}}\partial_{\mu}
\widetilde{P^{*\dag}_{\nu a}} V^{\mu}_{ba} \widetilde{P^{*\nu}_b}-
i\frac{\beta g_v}{\sqrt{2}}\widetilde{P^{*\dag}_{\nu a}}
 V^{\mu}_{ab}\partial_{\mu} \widetilde{P^{*\nu}_b}\nonumber\\
&-&i2\sqrt{2}\lambda g_v \bar{M^*}\widetilde{P^{*\mu\dag}_a}
(\partial_{\mu}V_{\nu}-\partial_{\nu}V_{\mu})_{ab}\widetilde{P^{*\nu}_b},\label{anti-vector-exchange}
\end{eqnarray}

\begin{eqnarray}
\mathcal{L}_{S}=-2g_s \bar{M}P_b\sigma P^{\dag}_{b}+ 2g_s
\bar{M^*}P^{*\mu}_b\sigma P^{*\dag}_{\mu b}\label{scalar-exchange}
\end{eqnarray}

\begin{eqnarray}
\widetilde{\mathcal{L}_{S}}=-2g_s
\bar{M}\widetilde{P^{\dag}_a}\sigma \widetilde{P_{a}}+ 2g_s
\bar{M^*}\widetilde{P^{*\dag}_{\mu a}}\sigma
\widetilde{P^{*\mu}_a}\label{anti-scalar-exchange}
\end{eqnarray}
where the heavy flavor meson fields $P$ and $P^*$ represent $P=(D^0,
D^+)$ or $(B^-, \bar{B}^0)$ and $P^*=(D^{*0}, D^{*+})$ or $(B^{*-},
\bar{B}^{*0})$. Its corresponding heavy anti-meson fields
$\widetilde{P}$ and $\widetilde{P}^*$ represent
$\widetilde{P}=(\bar{D}^0,D^-)$ or $(B^+, B^0)$ and
$\widetilde{P}^*=(\bar{D}^{*0},D^{*-})$ or $(B^{*+}, B^{*0})$.
$\phi$, $V$ represent the the exchanged pseudoscalar and vector
meson matrices. $\sigma$ is the only scalar meson interacting with
the heavy flavor meson.

\begin{eqnarray}
\phi=\left(
         \begin{array}{cc}
           \frac{\pi^0}{\sqrt{2}}+\frac{\eta}{\sqrt{6}} & \pi^+ \\
           \pi^- & -\frac{\pi^0}{\sqrt{2}}+\frac{\eta}{\sqrt{6}} \\
         \end{array}
       \right)
\end{eqnarray}

\begin{eqnarray}
V=\left(
         \begin{array}{cc}
           \frac{\rho^0}{\sqrt{2}}+\frac{\omega}{\sqrt{2}} & \rho^+ \\
           \rho^- & -\frac{\rho^0}{\sqrt{2}}+\frac{\omega}{\sqrt{2}} \\
         \end{array}
       \right)
\end{eqnarray}

According to the OBE model, five mesons ( $\pi$, $\sigma$, $\rho$,
$\omega$ and $\eta$) contribute to the effective potential. For the
$D^*\bar{D}^{*}$ and $B^*\bar{B}^{*}$ systems, the potentials are
the same for the three isovector states in Eqs. (1)$\sim$(4) with
the exact isospin symmetry. Expanding the Lagrangian densities in
Eqs. (5)$\sim$(10) leads to each meson's contribution for this
channel. These channel-dependent coefficients are listed in Table
\ref{tab:channel-coeff}.

The pionic coupling constant $g\!=\!0.59$ is extracted from the
width of $D^{*+}$\cite{S.Ahmed:2001}. $f_{\pi}$=132 MeV is the pion
decay constant. According the vector meson dominance mechanism, the
parameters $g_v$ and $\beta$ can be determined as $g_v=5.8$ and
$\beta=0.9$. At the same time, by matching the form factor obtained
from the light cone sum rule and that calculated from the lattice
QCD, we can get $\lambda=0.56$ GeV$^{-1}$ \cite{C.Isola:2003,
M.Bando:1988}. The coupling constant related to the scalar meson
exchange is $g_s=g_{\pi}/2\sqrt{6}$ with $g_{\pi}=3.73$
\cite{X.Liu:2009, A.F.Falk:1992}. All these parameters are listed in
Table \ref{tab:coupling-constant}.

\begin{table}[htbp]
\caption{coefficients } \label{tab:channel-coeff}
\begin{center}
\begin{tabular}{c| c  | c c c c c }
\hline \hline & \multirow{2}{*} {isospin}   & \multicolumn{5}{c}{meson-exchange }  \\
\cline{3-7}
& & $~~~\rho~~~$ & $~~~\omega~~~$ & $~~~\sigma~~~$ & $~~~\pi~~~$ & $~~~\eta~~~$\\
\hline
\multirow{2}{*}{$D^*\bar{D}^{*}$} & $I=1$ &   ~-1/2~ &  ~1/2~ & ~1~ & ~$-1/2$~   &  $1/6$  \\
\cline{2-7}
                                  & $I=0$ &   ~3/2~  & ~1/2~  & ~1~ & ~$3/2$~  &   $1/6$  \\
\hline
\multirow{2}{*}{$B^*\bar{B}^{*}$} & $I=1$ &   ~-1/2~ &  ~1/2~ & ~1~ & ~$-1/2$~   &  $1/6$  \\
\cline{2-7}
                                  & $I=0$ &   ~3/2~  & ~1/2~  & ~1~  & ~$3/2$~  &  $1/6$  \\
\hline\hline
\end{tabular}
\end{center}
\end{table}

\begin{table}[htbp]
\caption{The coupling constants and masses of the heavy mesons and
the exchanged light mesons used in our calculation. The masses of
the mesons are taken from the PDG \cite{PDG}}
\label{tab:coupling-constant}
\begin{center}
\begin{tabular}{c | c  | c }
\hline \hline  & {mass(MeV)}  & {coupling constants} \\
\cline{2-3} \hline
\multirow{2}{*}{pseudoscalar} & $m_{\pi}=134.98$  &  $g=0.59$  \\

                              &  $m_{\eta}=547.85$ &  $f_{\pi}=132 MeV$  \\
\hline
\multirow{3}{*}{vector}       & $m_{\rho}=775.49$       &  $g_v=5.8$  \\

                              & $m_{\omega}=782.65$     &  $\beta=0.9$  \\

                              &                         &  $\lambda=0.56 GeV^{-1}$  \\
\hline
\multirow{2}{*}{scalar}       & $m_{\sigma}=600$    &  $g_s=g_{\pi}/2\sqrt{6}$  \\

                              &   &  $g_{\pi}=3.73$  \\
\hline\hline
\multirow{2}{*}{heavy flavor }  &  $m_{D^*}=2010.25$     &   \\
                                &  $m_{B^*}=5325.0$     &   \\
\hline\hline
\end{tabular}
\end{center}
\end{table}

In order to include all the momentum-related terms in our
calculation, we introduce the polarization vector of the vector
mesons. At the rest frame we have
\begin{eqnarray}\label{pol}
\epsilon_{\lambda}=(0,\vec{\epsilon_{\lambda}})
\end{eqnarray}
We make a lorentz boost to Eq. \ref{pol} to derive the polarization
vector in the laboratory frame
\begin{eqnarray}
\epsilon^{lab}_{\lambda}=(\frac{\vec{p}\cdot\vec{\epsilon_{\lambda}}}{m},
\vec{\epsilon_{\lambda}}+\frac{\vec{p}
(\vec{p}\cdot\vec{\epsilon_{\lambda}})}{m(P_0 + m)})
\end{eqnarray}
where $p=(p_0,\mathbf{p})$ is the particle's 4-momentum in the
laboratory frame and $m$ is the mass of the particle.

\begin{center}
\textbf{B. Effective potential }
\end{center}

With the wave function and Feynman diagram, we can derive the
relativistic scattering amplitude at the tree level
\begin{equation}
\langle f | S | i \rangle = \delta_{fi} + i \langle f | T | i
\rangle = \delta_{fi} + (2\pi)^4\delta^4(p_f-p_i) i M_{fi},
\end{equation}
where the T-matrix is the interaction part of the S-matrix and
$M_{fi}$ is defined as the invariant matrix element. After applying
Bonn approximation to the Lippmann-Schwinger equation, the S-matrix
reads
\begin{equation}
\langle f | S | i \rangle = \delta_{fi} - 2\pi \delta(E_f-E_i) i
V_{fi}
\end{equation}
with $V_{fi}$ being the effective potential. Considering the
different normalization conventions used for the scattering
amplitude $M_{fi}$, $T$-matrix $T_{fi}$ and $V_{fi}$, we have
\begin{equation}
V_{fi}=-\frac{M_{fi}}{\sqrt{ \mathop\prod\limits_{f}2{p_f}^0
\mathop\prod\limits_{i} 2{p_i}^0}}\approx -\frac{M_{fi}}
{\sqrt{\mathop\prod\limits_{f} 2{m_f}^0 \mathop\prod\limits_{i}
2{m_i}^0}}
\end{equation}
where $p_{f(i)}$ denotes the four momentum of the final (initial)
state.

During our calculation, $P_1=(E_1,\vec{p})$ and $P_2=(E_2,-\vec{p})$
denote the four momenta of the initial states in the center mass
system, while $P_3=(E_3,\vec{p'})$ and $P_4=(E_4,-\vec{p'})$ denote
the four momenta of the final states, respectively.
\begin{equation}
q=P_3-P_1=(E_3-E_1,\vec{p'}-\vec{p})=(E_2-E_4,\vec{q})
\end{equation}
is the transferred four momentum or the four momentum of the meson
propagator. For convenience, we always use
\begin{equation}
\vec{q}=\vec{p'}-\vec{p}
\end{equation}
and
\begin{equation}
\vec{k}=\frac{1}{2}(\vec{p'}+\vec{p})
\end{equation}
instead of $\vec{p'}$ and $\vec{p}$ in the practical calculation.

In the OBE model, a form factor is introduced at each vertex to
suppress the high momentum contribution. We take the conventional
form for the form factor as in the Bonn potential model.
\begin{equation}
F(q)=\frac{\Lambda^2-m_{\alpha}^2}{\Lambda^2-q^2}=\frac{\Lambda^2-m_{\alpha}^2}{{\Lambda}^2+\vec{q}^2}
\end{equation}
$m_\alpha$ is the mass of the exchanged meson and $m^*$ is the mass
of the heavy flavor meson $D^*$ or $B^*$. So far, the effective
potential is derived in the momentum space. In order to solve the
time independent Schr\"{o}dinger equation in the coordinate space,
we need to make the Fourier transformation to $V(\vec{q},\vec{k})$.
The details of the Fourier transformations are presented in the
Appendix.

The expressions of the potential through exchanging the $\sigma$,
$\rho$ mesons are
\begin{eqnarray}
V_{\sigma}&=&-C_{\sigma}g^2_s(\vec{\epsilon_b}\cdot
\vec{\epsilon_a}^{\dag})(\vec{\epsilon_{b'}}\cdot
\vec{\epsilon_{a'}}^{\dag})F_{1t\sigma}\nonumber\\
&~&-C_{\sigma}g^2_s \frac{1}{m^{*2}}[(\vec{\epsilon_b}\cdot
\vec{\epsilon_a}^{\dag})(\vec{\epsilon_{b'}}\cdot
\vec{\epsilon_{a'}}^{\dag})F_{3t1\sigma}+S^{'}_{12}F_{3t2\sigma}]\nonumber\\
&~&+C_{\sigma}g^2_s \frac{1}{m^{*2}}i(\vec{\epsilon_{b'}}\cdot
\vec{\epsilon_{a'}}^{\dag})[(\vec{\epsilon_a}^{\dag}\times
\vec{\epsilon_b})\cdot \vec{L}]F_{5t\sigma}
\end{eqnarray}

\begin{eqnarray}
V_{\rho}&=&-C_{\rho}\frac{\beta^2 g^2_v}{2} (\vec{\epsilon_b}\cdot
\vec{\epsilon_a}^{\dag})(\vec{\epsilon_{b'}}\cdot
\vec{\epsilon_{a'}}^{\dag})F_{1t\rho}\nonumber\\
&~&-C_{\rho}2\lambda^2 g^2_v (\vec{\epsilon_b}\times
\vec{\epsilon_a}^{\dag})(\vec{\epsilon_{b'}}\times
\vec{\epsilon_{a'}}^{\dag})F_{2t\rho}\nonumber\\
&~&+C_{\rho}2\lambda^2 g^2_v [(\vec{\epsilon_b}\times
\vec{\epsilon_a}^{\dag})(\vec{\epsilon_{b'}}\times
\vec{\epsilon_{a'}}^{\dag})F_{3t1\rho}+ \widetilde{S}_{12}F_{3t2\rho}]\nonumber\\
&~&-C_{\rho}(\frac{\beta^2 g^2_v}{2m^{*2}}-\frac{2\lambda \beta
g^2_v}{m^*} )[(\vec{\epsilon_b}\cdot
\vec{\epsilon_a}^{\dag})(\vec{\epsilon_{b'}}\cdot
\vec{\epsilon_{a'}}^{\dag})F_{3t1\rho}+ S^{'}_{12}F_{3t2\rho}]\nonumber\\
&~&-C_{\rho} \frac{\beta^2 g^2_v}{2 m^{*2}}(\vec{\epsilon_b}\cdot
\vec{\epsilon_a}^{\dag})(\vec{\epsilon_{b'}}\cdot
\vec{\epsilon_{a'}}^{\dag})[F_{4t1\rho}+\{-\frac{1}{2}\nabla^2, F_{4t2\rho}\}]\nonumber\\
&~&+C_{\rho}( \frac{\beta^2 g^2_v }{2m^{*2}}-\frac{4\lambda \beta
g^2_v}{m^*})i(\vec{\epsilon_{b'}}\cdot
\vec{\epsilon_{a'}}^{\dag})[(\vec{\epsilon_a}^{\dag}\times
\vec{\epsilon_b})\cdot \vec{L}]F_{5t\rho}
\end{eqnarray}
The $\omega$ and $\rho$ meson exchange potentials have the same form
except that the meson mass and channel-dependent coefficients are
different.

The expression of the potential through exchanging the $\pi$ meson
is
\begin{eqnarray}
V_{\pi}&=& C_{\pi}
\frac{g^2_{\pi}}{f^2_{\pi}}[(\vec{\epsilon_b}\times
\vec{\epsilon_a}^{\dag})(\vec{\epsilon_{b'}}\times
\vec{\epsilon_{a'}}^{\dag})F_{3t1\pi}+ \widetilde{S}_{12}F_{3t2\pi}]
\end{eqnarray}
where $S^{'}_{12}$ and $\widetilde{S}_{12}$ have the form
\begin{eqnarray}
S^{'}_{12}=[3(\vec{r}\cdot\vec{\epsilon_b})(\vec{r}\cdot\vec{\epsilon_a}^{\dag})-(\vec{\epsilon_b}\cdot
\vec{\epsilon_a}^{\dag})](\vec{\epsilon_{b'}}\cdot
\vec{\epsilon_{a'}}^{\dag})
\end{eqnarray}

\begin{eqnarray}
\widetilde{S}_{12}=3[\vec{r}\cdot(\vec{\epsilon_b}\times
\vec{\epsilon_a}^{\dag})][\vec{r}\times(\vec{\epsilon_{b'}}\cdot
\vec{\epsilon_{a'}}^{\dag})]-(\vec{\epsilon_b}\times
\vec{\epsilon_a}^{\dag})(\vec{\epsilon_{b'}}\times
\vec{\epsilon_{a'}}^{\dag})
\end{eqnarray}

Compared to the $D \bar{D}^*$ case, there appear several new
interaction operators: $(\vec{\epsilon_b}\times
\vec{\epsilon_a}^{\dag})(\vec{\epsilon_{b'}}\times
\vec{\epsilon_{a'}}^{\dag})$, $S^{'}_{12}$, $\widetilde{S}_{12}$ and
$i(\vec{\epsilon_{b'}}\cdot\vec{\epsilon_{a'}}^{\dag})[(\vec{\epsilon_a}^{\dag}\times
\vec{\epsilon_b})\cdot \vec{L}]$. These operator represent the new
form of the tensor, spin-spin and spin-orbit interactions.

Similarly, the $\eta$ and $\pi$ meson exchange potential has the
same form in the $D^* {\bar D}^{*}$ and $B^* {\bar B}^{*}$ system
except the meson mass and channel-dependent coefficients.
The explicit forms of
$\mathcal{F}_{\mu t \alpha}$,$\mathcal{F}_{\mu u
 \alpha}$,$\mathcal{F}_{\mu t \nu \alpha}$, $\mathcal{F}_{\mu u
\nu \alpha}$ are shown in the Appendix.

In our calculation, we explicitly consider the external momentum of
the initial and final states. Due to the recoil corrections, several
new terms appear which were omitted in the heavy quark symmetry
limit. These momentum dependent terms are related to the momentum
$\vec{k}=\frac{1}{2}(\vec{p'}+\vec{p})$:
\begin{eqnarray}
\frac{\vec{k}^2}{\vec{q}^2+m_{\alpha}^2}
\end{eqnarray}
and
\begin{eqnarray}\label{so}
{\frac{i(\vec{\epsilon_{b'}}\cdot
\vec{\epsilon_{a'}}^{\dag})[(\vec{\epsilon_a}^{\dag}\times
\vec{\epsilon_b})\cdot(\vec{k}\times\vec{q})]}{\vec{q}^2+m_{\alpha}^2}}
\end{eqnarray}

The term in Eq. (\ref{so}) is the well-known spin orbit force. In
short, all the terms in the effective potentials in the form of
$F_{3t1\rho}$, $F_{4t1\rho}$, $F_{5t\rho}$ etc with the sub-indices
$3, 4, 5$ arise from the recoil corrections and vanish when the
heavy meson mass $m^*$ goes to infinity. The recoil correction and the spin
orbit force appear at $O(1/M^2)$.

\begin{center}
\textbf{C.  Schr\"{o}dinger equation }
\end{center}
With the effective potential $V(\vec{r})$ in Eqs. (23) $\sim$ (27),
we are able to study the binding property of the system by solving
the Schr\"{o}dinger Equation
\begin{equation}
(-\frac{\hbar^2}{2\mu}\nabla^{2}+V(\vec{r})-E)\Psi(\vec{r})=0,
\label{eq:schrod}
\end{equation}
where $\Psi(\vec{r})$ is the total wave function of the system. The
total spin of the system $S=1$ and the orbital angular momenta $L=0$
and $L=2$. Thus the wave function $\Psi(\vec{r})$ should have the
following form
\begin{equation}
\Psi(\vec{r})=\psi_S(\vec{r})+\psi_D(\vec{r}),
\end{equation}
where $\psi_S(\vec{r})$ and $\psi_D(\vec{r})$ are the $S$-wave and
$D$-wave functions, respectively. We use the same matrix method in
Ref. \cite{L.Zhao:2013} to solve this S-D wave couple-channel
equation.

We detach the terms related to the
kinetic-energy-operator $\nabla^{2}$ from $V(\vec{r})$
and re-write Eq. (\ref{eq:schrod}) as
\begin{eqnarray}
&~&(-\frac{\hbar^2}{2\mu}\nabla^{2}-\frac{\hbar^2}{2\mu}[\nabla^2
\alpha(r)+\alpha(r)\nabla^2]\nonumber\\
&~&+\widetilde{V}(\vec{r})-E~)\Psi(\vec{r})=0
\end{eqnarray}
with
\begin{equation}
\nabla^2=\frac{1}{r}\frac{d^2}{dr^2}r-\frac{\overrightarrow{L}^2}{r^2},
\end{equation}
in which $\alpha(r)$ is
\begin{eqnarray}
\alpha(r)&=&(-2\mu)[-C_{\rho} \frac{\beta^2 g^2_v}{2
m^{*2}}(\vec{\epsilon_b}\cdot
\vec{\epsilon_a}^{\dag})(\vec{\epsilon_{b'}}\cdot
\vec{\epsilon_{a'}}^{\dag}) \mathcal{F}_{4t2 \rho}\nonumber\\
&~&-C_{\rho} \frac{\beta^2 g^2_v}{2 m^{*2}}(\vec{\epsilon_b}\cdot
\vec{\epsilon_a}^{\dag})(\vec{\epsilon_{b'}}\cdot
\vec{\epsilon_{a'}}^{\dag}) \mathcal{F}_{4t2\omega}]
\end{eqnarray}

The total Hamiltonian contains three angular momentum related
operators $(\vec{\epsilon_b}\cdot
\vec{\epsilon_a}^{\dag})(\vec{\epsilon_{b'}}\cdot
\vec{\epsilon_{a'}}^{\dag})$, $S^{'}_{12}$, $(\vec{\epsilon_b}\times
\vec{\epsilon_a}^{\dag})(\vec{\epsilon_{b'}}\times
\vec{\epsilon_{a'}}^{\dag})$, $\widetilde{S}_{12}$,
$i(\vec{\epsilon_{b'}}\cdot
\vec{\epsilon_{a'}}^{\dag})[(\vec{\epsilon_a}^{\dag}\times
\vec{\epsilon_b})\cdot \vec{L}]$, which corresponds to the spin-spin
interaction, spin orbit force and tensor force respectively. They
act on the S and D-wave coupled wave functions $\phi_S+\phi_D$ and
split the total effective potential $\widetilde{V}(\vec{r})$ into
the subpotentials $V_{SS}(r)$, $V_{SD}(r)$, $V_{DS}(r)$ and
$V_{DD}(r)$. The matrix form reads
\begin{eqnarray}
\langle \phi_S+\phi_D|\!\!\!\!\!\!\!&~&(\vec{\epsilon_b}\cdot
\vec{\epsilon_a}^{\dag})(\vec{\epsilon_{b'}}\cdot
\vec{\epsilon_{a'}}^{\dag})\widetilde{V}(\vec{r})~|\phi_S+\phi_D\rangle\nonumber\\
&~&=\left(\begin{array}{cc}
V_{SS}(r) & 0 \\
0 & V_{DD}(r) \\
\end{array}
\right)
\end{eqnarray}

\begin{eqnarray}
\langle \phi_S+\phi_D |S^{'}_{12}\widetilde{V}(\vec{r})
|\phi_S+\phi_D\rangle=\left(\!
         \begin{array}{cc}
           \!\!0 & \!\!\frac{1}{\sqrt{2}}V_{SD}(r) \\
           \!\!\frac{1}{\sqrt{2}}V_{DS}(r) & \!\!-\frac{1}{2} \\
         \end{array}
      \! \right)
\end{eqnarray}

\begin{eqnarray}
\langle \phi_S+\phi_D|\!\!\!\!\!\!\!&~&(\vec{\epsilon_b}\times
\vec{\epsilon_a}^{\dag})(\vec{\epsilon_{b'}}\times
\vec{\epsilon_{a'}}^{\dag})\widetilde{V}(\vec{r})~|\phi_S+\phi_D\rangle\nonumber\\
&~&=\left(\begin{array}{cc}
V_{SS}(r) & 0 \\
0 & V_{DD}(r) \\
\end{array}
\right)
\end{eqnarray}

\begin{eqnarray}
\langle \phi_S+\phi_D |\widetilde{S}_{12}\widetilde{V}(\vec{r})
|\phi_S+\phi_D\rangle=\left(\!
         \begin{array}{cc}
           \!\!0 & \!\!-\sqrt{2}V_{SD}(r) \\
           \!\!-\sqrt{2}V_{DS}(r) & \!\!1 \\
         \end{array}
      \! \right)
\end{eqnarray}

\begin{eqnarray}
\langle \phi_S+\phi_D|\!\!\!\!\!\!\!&~&i(\vec{\epsilon_{b'}}\cdot
\vec{\epsilon_{a'}}^{\dag})[(\vec{\epsilon_a}^{\dag}\times
\vec{\epsilon_b})\cdot \vec{L}]\widetilde{V}(\vec{r})~|\phi_S+\phi_D\rangle\nonumber\\
&~&=\left(\begin{array}{cc}
0 & 0 \\
0 & \frac{3}{2}V_{DD}(r) \\
\end{array}
\right)
\end{eqnarray}

\section{Numerical Results for the $B^* \bar{B}^{*}$ system}\label{Numerical-BB}

We diagonalize the Hamiltonian matrix to obtain the eigenvalue and
eigenvector. If there exists a negative eigenvalue, there exists a
bound state. The corresponding eigenvector is the wave function. We
use the variation principle to solve the equation. We change the
variable parameter to get the lowest eigenvalue. We also change the
number of the basis functions to reach a stable result.

\subsection{$Z_b(10650)$ }

\begin{figure}[ht]
  \begin{center}
  \rotatebox{0}{\includegraphics*[width=0.38\textwidth]{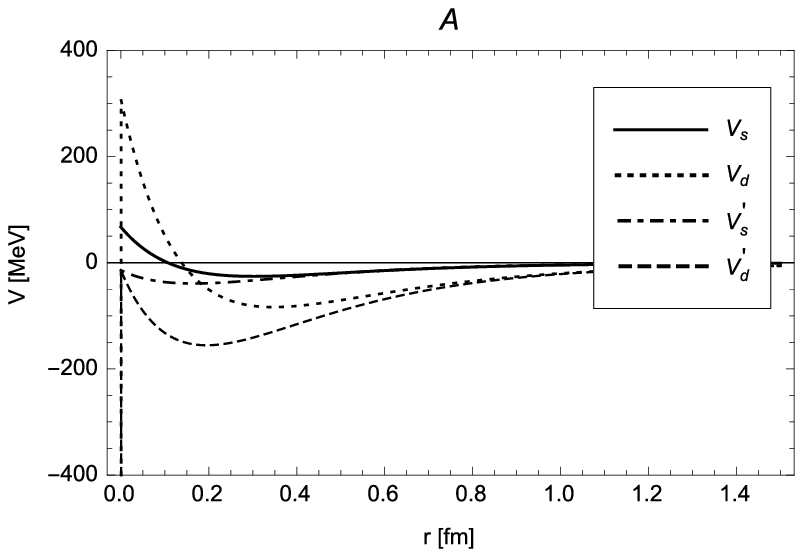}}
    \rotatebox{0}{\includegraphics*[width=0.38\textwidth]{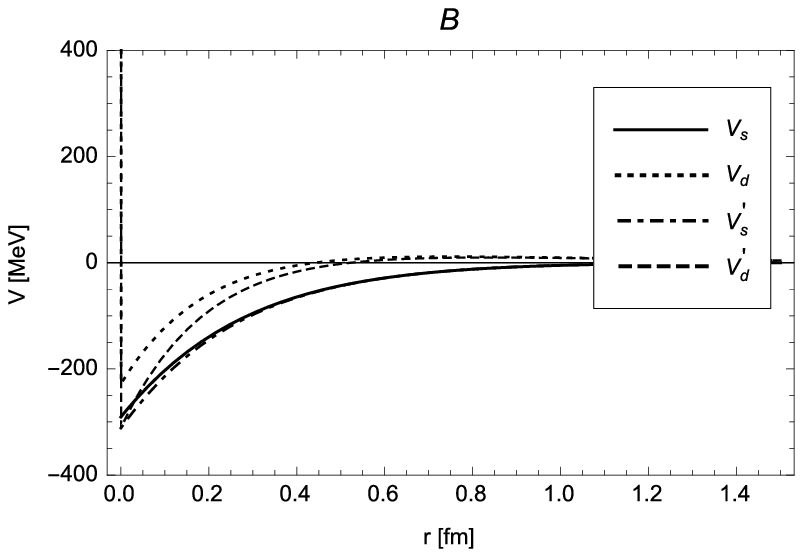}}
      \rotatebox{0}{\includegraphics*[width=0.38\textwidth]{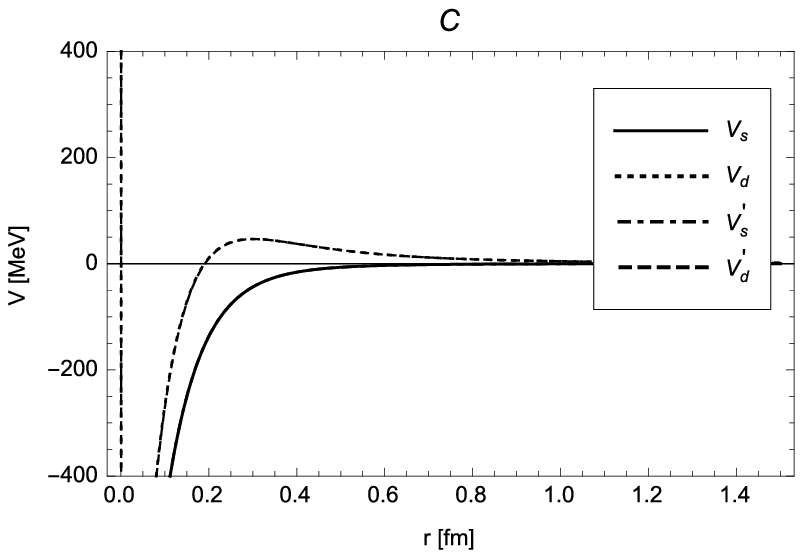}}
    \rotatebox{0}{\includegraphics*[width=0.38\textwidth]{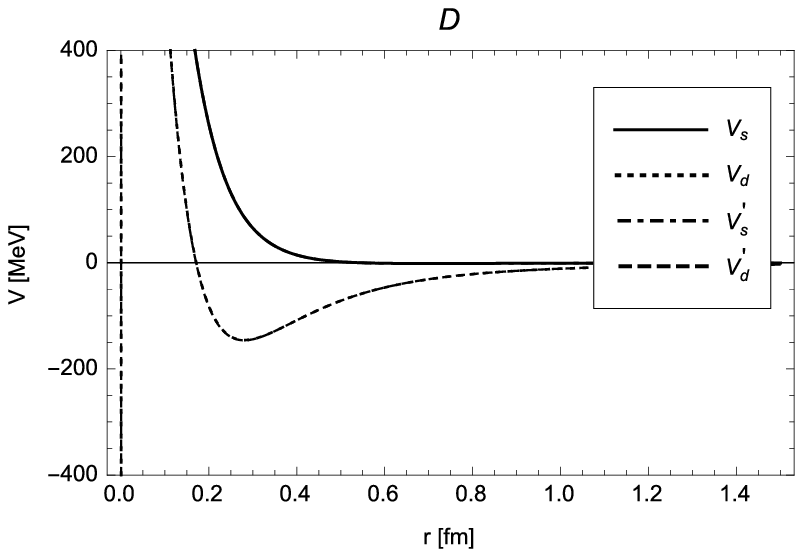}}
    \caption{The effective potential of the $B^* \bar{B}^{*}$ system.
    The Labels $A$,$B$,$C$, and $D$ correspond to the four cases $I=0^-$, $J^{PC}=1^{+-}$; $I=0^+$, $J^{PC}=1^{++}$;
    $I=1^+$, $J^{PC}=1^{+-}$; and $I=1^-$, $J^{PC}=1^{++}$, respectively
    from top to bottom. $V_s$ and $V_d$ are the effective potentials of the $S$-wave and $D$-wave
    interactions with the momentum-related terms, while $V'_s$ and $V'_d$ are
    the $S$-wave and $D$-wave effective potentials without the momentum-related terms. }
    \label{B-potential}
  \end{center}
\end{figure}

Since the mass of the charged bottomonium-like state $Z_b(10650)$ is
close to the $B^* \bar{B}^{*}$ system, we first consider the
possibility of the $B^* \bar{B}^{*}$ molecule with $I^G=1^+$,
$J^{PC}=1^{+-}$. In order to reflect the recoil correction of the
momentum-related terms, we plot the effective potential of the
S-wave and D-wave with or without the momentum-related terms in Fig.
\ref{B-potential}. $V_s$ and $V_d$ are the effective potentials of
the $S$-wave and $D$-wave interactions after adding the
momentum-related terms. $V'_s$ and $V'_d$ are the effective
potentials of the $S$-wave and $D$-wave interactions without the
momentum-related terms. Fig. \ref{B-potential} C corresponds to the
$I^G=1^+$, $J^{PC}=1^{+-}$ $B^* \bar{B}^{*}$ system, where the
curves of $V_s$ and $V'_s$, and $V_d$ and $V'_d$ are almost
overlapping. In other words, the recoil correction is small.

We collect the numerical results in Table \ref{tab:BB1+}. $E$ and
$E'$ are the eigenenergy of Hamiltonian with and without the
momentum-related terms, respectively. The fourth, fifth and sixth
column represent the contribution of $S$-wave, $D$-wave, and
spin-orbit force components, respectively. The last column is the
mass of $B^* \bar{B}^{*}$ as a molecular state of $I^G=1^+$,
$J^{PC}=1^{+-}$. When the cut off lies within $2.2-2.8$ GeV, there
exists a bound-state. The binding energy with the recoil correction
is between $0.97-15.15$ MeV. The binding energy without the recoil
correction is between $0.94-14.98$ MeV. When the cutoff parameter
$\Lambda = 2.2$ GeV, the binding energy is $0.97$ MeV, and the
recoil correction is only $-0.03$ MeV. The contribution from the
spin-orbit force is as small as $0.001$ MeV. When the cutoff
parameter $\Lambda = 2.8$ GeV, the binding energy is $15.15$ MeV,
and the recoil correction is $-0.07$ MeV. The correspondence
spin-orbit force contribution is $0.02$ MeV, which is also small
compared with the binding energy. The recoil correction and the
contribution of the spin-orbit are very small. However the recoil
correction is favorable to the formation of the $B^* \bar{B}^{*}$
molecular state with $I^G=1^+$, $J^{PC}=1^{+-}$.

\begin{table}[htbp]
\caption{The bound state solution of the $B^* \bar{B}^{*}$ system
with $I^G=1^+$, $J^{PC}=1^{+-}$ (in unit of MeV) and different
$\Lambda$. $E$ and $E'$ is the eigen-energy of the system with and
without the momentum-related terms respectively. We also list the
separate contribution to the energy from the S-wave, D-wave and
spin-orbit force components respectively in the fourth, fifth and
sixth column. The last column is the mass of the $B^* \bar{B}^{*}$
system with $I^G=1^+$, $J^{PC}=1^{+-}$ as a molecular state.}
\label{tab:BB1+}
\begin{center}
\begin{tabular}{c c c c c c | c }
\hline \hline \multirow{2}{*}{~$\Lambda$(GeV)~} & \multirow{2}{*}{~}  &\multicolumn{4}{c|}{Eigenvalue} & {Mass}   \\
\cline{2-6}
                                &   &total  &   S           &D              & LS          &(MeV) \\
\hline
\multirow{2}{*}{2.2} & $E$    &  ~~-0.97~~  & ~~13.15~~ & ~~0.15~~   & ~~0.001~~  & ~~10649.03~~ \\
                               & $E'$   &  ~~-0.94~~  & ~~-12.91~~ & ~~-0.15~~   & ~~-~~  & ~~10649.06~~\\
\hline\hline
\multirow{2}{*}{2.4} & $E$    &  ~~-3.49~~  & ~~-28.68~~ &~~0.32~~    & ~~0.004~~   & ~~10646.51~~\\
                               & $E'$   &  ~~-3.43~~  & ~~-28.27~~ &~~0.32~~    & ~~-~~ & ~~10646.57~~\\
\hline\hline
\multirow{2}{*}{2.6} & $E$    &  ~~-8.04~~  & ~~-49.81~~ &~~0.56~~    & ~~0.01~~  & ~~10641.96~~\\
                               & $E'$   &  ~~-7.94~~  & ~~-49.19~~ & ~~0.55~~   & ~~-~~ & ~~10642.06~~\\
\hline\hline
\multirow{2}{*}{2.8} & $E$    &  ~~-15.15~~  & ~~-77.53~~ &~~0.88~~   & ~~0.02~~  & ~~10634.85~~\\
                               & $E'$   &  ~~-14.98~~   & ~~-76.66~~ &~~0.87~~   & ~~-~~  & ~~10635.02~~\\
\hline\hline
\end{tabular}
\end{center}
\end{table}

\begin{figure}[ht]
  \begin{center}
  \rotatebox{0}{\includegraphics*[width=0.38\textwidth]{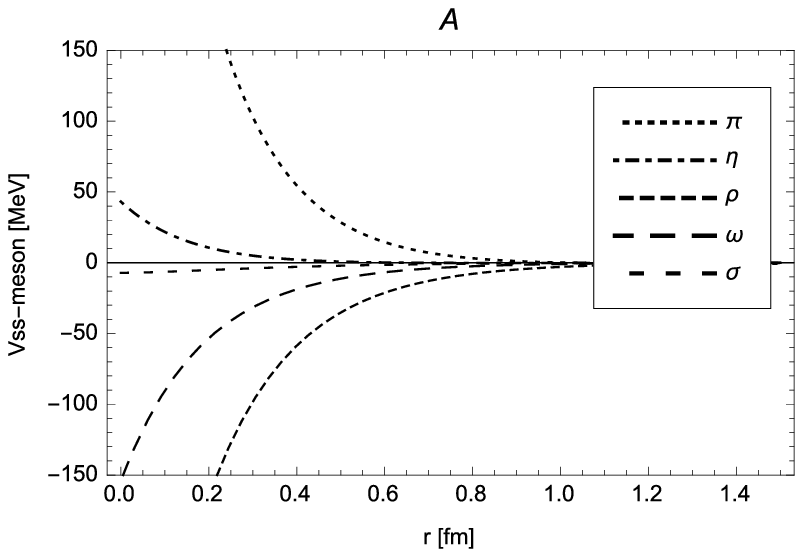}}
    \rotatebox{0}{\includegraphics*[width=0.38\textwidth]{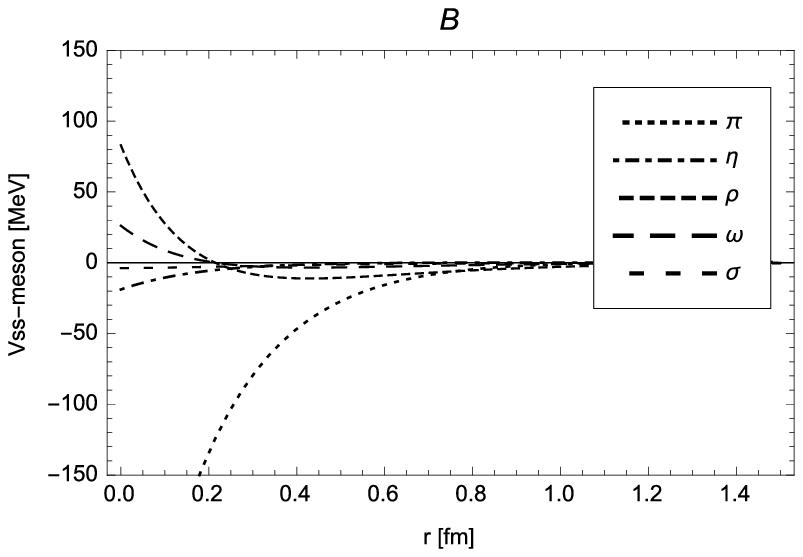}}
      \rotatebox{0}{\includegraphics*[width=0.38\textwidth]{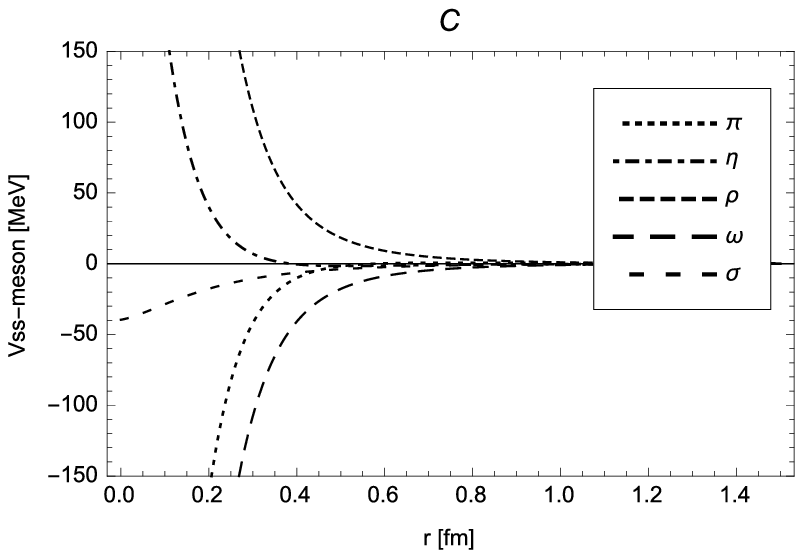}}
    \rotatebox{0}{\includegraphics*[width=0.38\textwidth]{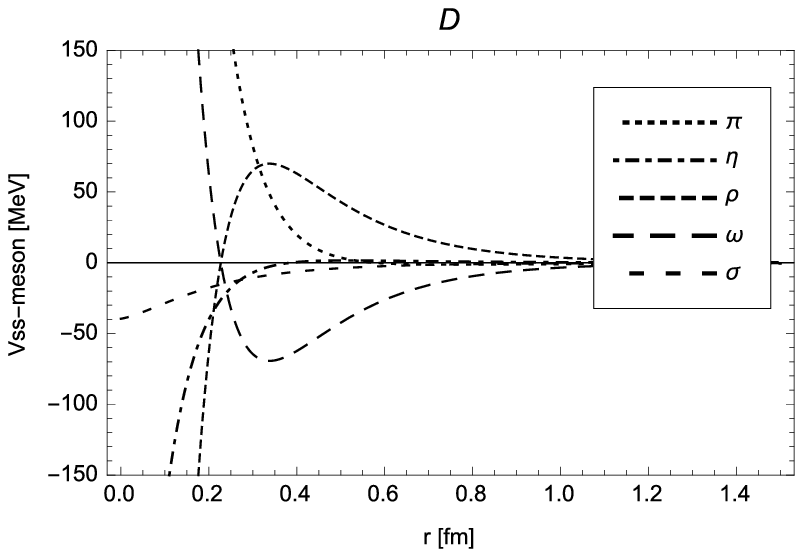}}
    \caption{The effective potential from the different meson exchange
    in the $B^* \bar{B}^{*}$ system. Labels A,B,C,D are the same as in
    Fig. \ref{B-potential}.}
    \label{B-potential-meson}
  \end{center}
\end{figure}

From Fig \ref{B-potential-meson} C, it is clear that the $\pi$
exchange is much more important than the other meson-exchanges.
Considering that the coupling constant $g$ is extracted from the
$D^*$ decay width with some uncertainty, we multiply $g$ by a factor
from $0.99$ to $1.1$ to check the dependence of the binding energy
on this parameter. The numerical results are listed in Table
\ref{tab:BB1+g}. The binding energy with the recoil correction
varies from $6.39-36.81$ MeV. The binding energy without the recoil
correction varies from $6.3-36.57$ MeV. The binding energy is
sensitive to the coupling constant.

\begin{table}[htbp]
\caption{The $B^* \bar{B}^{*}$ system with $I^G=1^+$,
$J^{PC}=1^{+-}$ (in units of MeV) and different coupling constant
$g$ and $\Lambda =2.0$ GeV. The other notations are the same as in
Table \ref{tab:BB1+}.} \label{tab:BB1+g}
\begin{center}
\begin{tabular}{c c c c c c | c}
\hline \hline \multirow{2}{*}{~$\Lambda $(GeV)~} & \multirow{2}{*}{~}  &\multicolumn{4}{c|}{Eigenvalue} & {Mass}   \\
\cline{2-6}
                                 &   &$Total$  &   $S$  &$D$   & $LS$ &(MeV) \\
\hline\hline
\multirow{2}{*}{$g\cdot0.99$}  & $E$    &  ~~$-6.39$~~  & ~~$-43.28$~~ & ~~$0.49$~~ & ~~$0.01$~~& ~~$10643.61$~~\\
                                               & $E'$   &  ~~$-6.30$~~  & ~~$-42.70$~~ &~~$0.49$~~ & ~~-~~& ~~$10643.7$~~\\
\hline\hline
\multirow{2}{*}{$g$}           & $E$    &  ~~$-8.04$~~  & ~~$-49.81$~~ &~~$0.56$~~ & ~~$0.01$~~& ~~$10641.96$~~\\
                                          & $E'$   & ~~$-7.94$~~   & ~~$-49.19$~~ &~~$0.55$~~ & ~~-~~& ~~$10642.06$~~\\
\hline\hline
\multirow{2}{*}{$g\cdot1.01$}  & $E$    &  ~~$-9.91$~~  & ~~$-56.68$~~ &~~$0.63$~~ & ~~$0.01$~~& ~~$10640.09$~~\\
                                                & $E'$   &  ~~$-9.79$~~  & ~~$-56.02$~~ &~~$0.63$~~ & ~~-~~    & ~~$10640.21$~~\\
\hline\hline
\multirow{2}{*}{$g\cdot1.1$}   & $E$    &  ~~$-36.81$~~  & ~~$-132.98$~~ &~~$1.32$~~ & ~~$0.02$~~& ~~$10613.19$~~\\
                                               & $E'$   &  ~~$-36.57$~~  & ~~$-131.96$~~ &~~$1.31$~~ & ~~-~~   & ~~$10613.43$~~\\
\hline\hline
\end{tabular}
\end{center}
\end{table}

\begin{figure}[ht]
  \begin{center}
  \rotatebox{0}{\includegraphics*[width=0.38\textwidth]{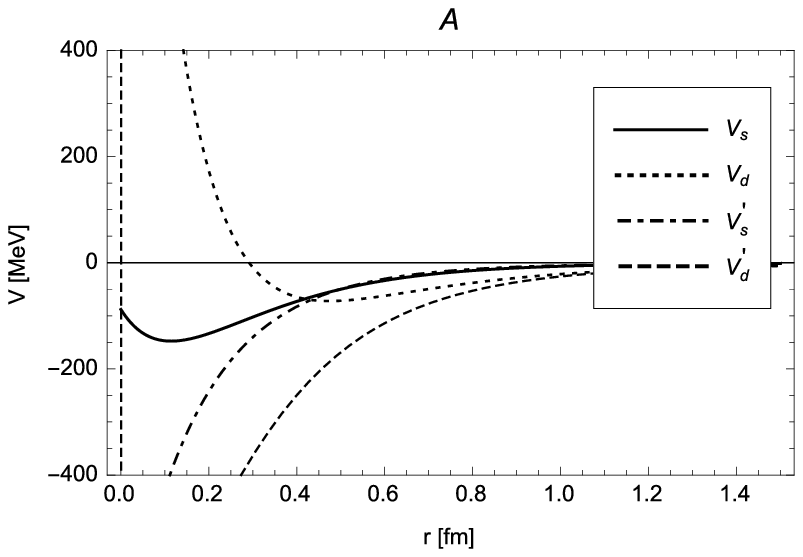}}
    \rotatebox{0}{\includegraphics*[width=0.38\textwidth]{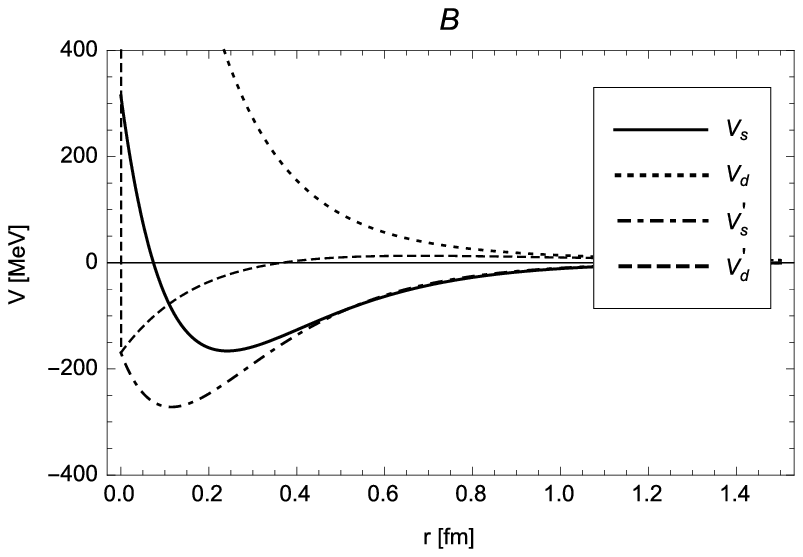}}
      \rotatebox{0}{\includegraphics*[width=0.38\textwidth]{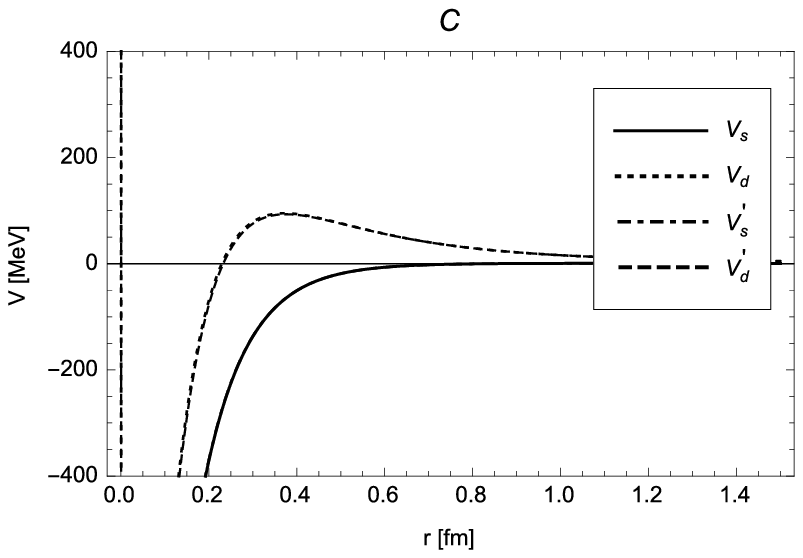}}
    \rotatebox{0}{\includegraphics*[width=0.38\textwidth]{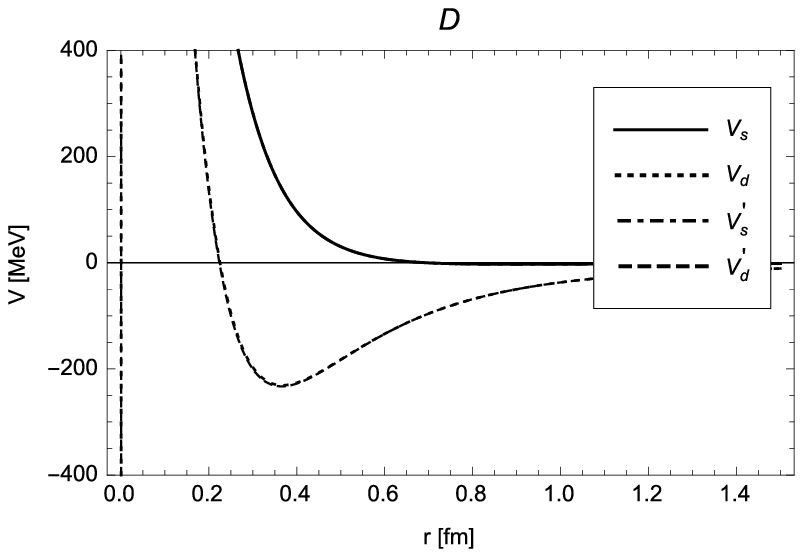}}
    \caption{The effective potential of the $D^* \bar{D}^{*}$ system.
    The Labels $A$,$B$,$C$, and $D$ correspond to the four cases $I=0^-$, $J^{PC}=1^{+-}$; $I=0^+$, $J^{PC}=1^{++}$;
    $I=1^+$, $J^{PC}=1^{+-}$; and $I=1^-$, $J^{PC}=1^{++}$, respectively
    from top to bottom. $V_s$ and $V_d$ are the effective potentials of the $S$-wave and $D$-wave
    interactions with the momentum-related terms, while $V'_s$ and $V'_d$ are
    the $S$-wave and $D$-wave effective potentials without the momentum-related terms. }
    \label{D-potential}
  \end{center}
\end{figure}

\begin{figure}[ht]
  \begin{center}
  \rotatebox{0}{\includegraphics*[width=0.38\textwidth]{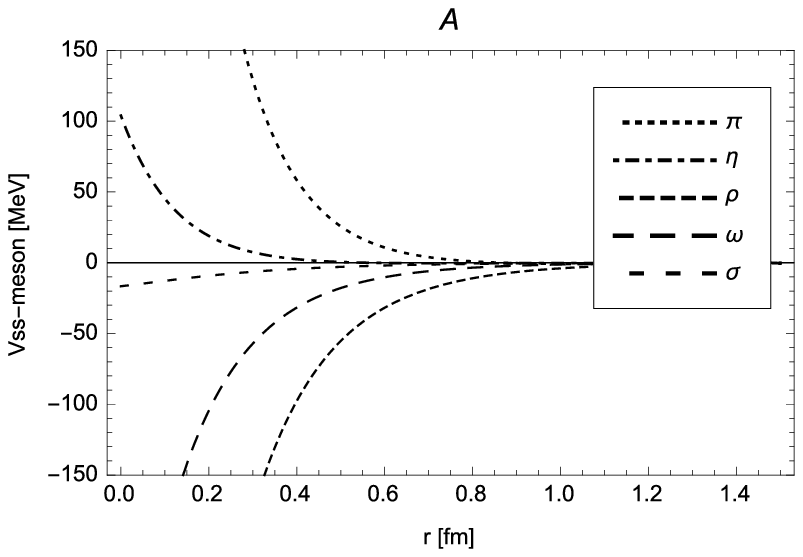}}
    \rotatebox{0}{\includegraphics*[width=0.38\textwidth]{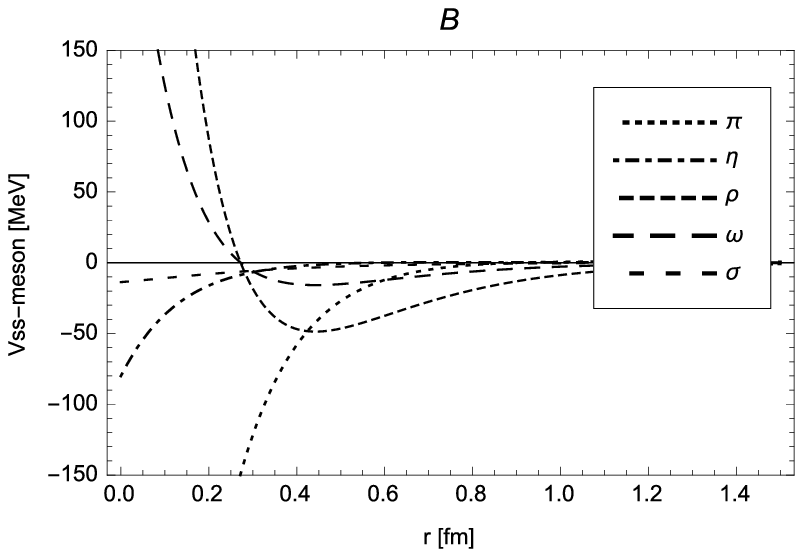}}
      \rotatebox{0}{\includegraphics*[width=0.38\textwidth]{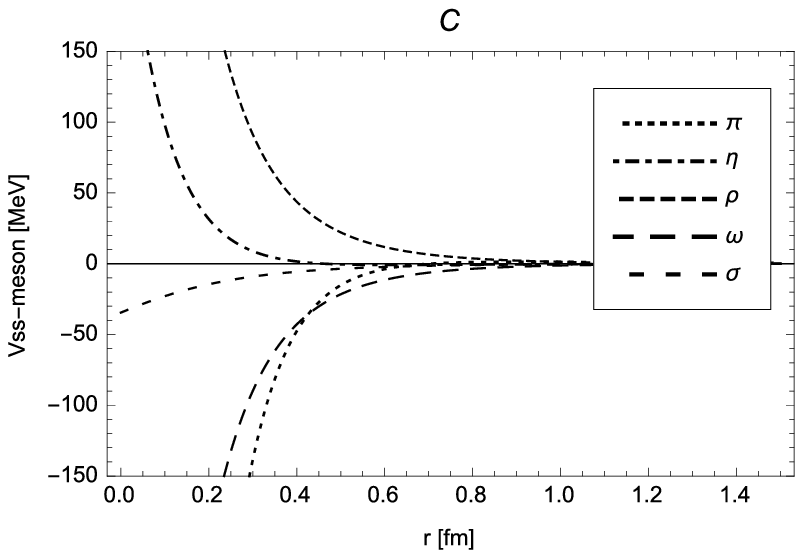}}
    \rotatebox{0}{\includegraphics*[width=0.38\textwidth]{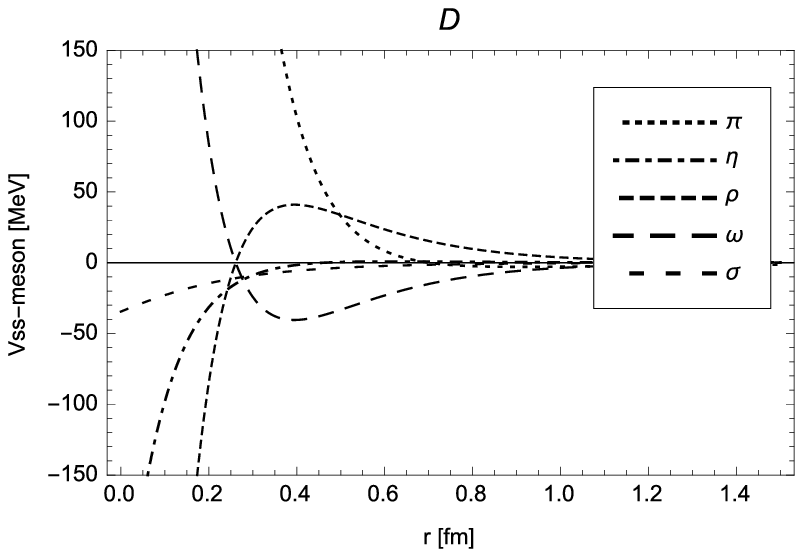}}
    \caption{The effective potential from the different meson exchange
    in the $D^* \bar{D}^{*}$ system. Labels A,B,C,D are the same as in
    Fig. \ref{D-potential}.}
    \label{D-potential-meson}
  \end{center}
\end{figure}

\subsection{The $B^*\bar{B}^*$ system with $I^G=1^-$, $J^{PC}=1^{++}$ }

For the $B^*\bar{B}^*$ system with $I^G=1^-$, $J^{PC}=1^{++}$, there
is no bound state when the cutoff varies in a reasonable range. We
multiply the coupling constant $g$ by a factor to investigate the
dependence of $g$. We list the results in Table \ref{tab:BB1-}. The
cutoff parameter is fixed at $\Lambda=2.0$ GeV. There appears a
bound state when $g$ increases by a factor $1.3$. When the factor
changes from $1.3-1.5$, the binding energy with recoil correction is
between $0.82-19.25$ MeV. The binding energy without the recoil
correction is between $0.8-19.18$ MeV. When the binding energy is
$0.82$, the recoil correction is $-0.02$ MeV, and the contribution
of the spin-orbit is $0.004$ MeV. When the binding energy is $19.25$
MeV, the the recoil correction is $-0.07$ MeV, and the contribution
of the spin-orbit force is $0.04$ MeV. From Fig \ref{B-potential} D,
the effect of the recoil correction is very small for the
$B^*\bar{B}^*$ system with $I^G=1^-$, $J^{PC}=1^{++}$. But it is
favorable for the formation of the molecular state. Fig
\ref{B-potential-meson} D shows the contributions of each
meson-exchange.

\begin{table}[htbp]
\caption{The $B^* \bar{B}^{*}$ system with $I^G=1^-$,
$J^{PC}=1^{++}$ (in units of MeV) with the variation of the coupling
constant $g$ and $\Lambda =2.0$ GeV. The other notations are the
same as in Table \ref{tab:BB1+}.} \label{tab:BB1-}
\begin{center}
\begin{tabular}{c c c c c c | c}
\hline \hline \multirow{2}{*}{~$\Lambda $(GeV)~} & \multirow{2}{*}{~}  &\multicolumn{4}{c|}{Eigenvalue} & {Mass}   \\
\cline{2-6}
                                 &   &$Total$  &   $S$  &$D$   & $LS$ &(MeV) \\
\hline\hline
\multirow{2}{*}{$g\cdot1.3$}  & $E$    &  ~~$-0.82$~~  & ~~$1.99$~~ & ~~$-2.23$~~ & ~~$0.004$~~& ~~$10649.18$~~\\
                                               & $E'$   &  ~~$-0.80$~~  & ~~$1.97$~~ &~~$-2.21$~~ & ~~-~~& ~~$10649.2$~~\\
\hline\hline
\multirow{2}{*}{$g\cdot1.4$}   & $E$    &  ~~$-6.68$~~  & ~~$7.73$~~ &~~$-8.94$~~ & ~~$0.02$~~& ~~$10643.32$~~\\
                                               & $E'$   &  ~~$-6.63$~~  & ~~$7.69$~~ &~~$-8.92$~~ & ~~-~~   & ~~$10643.37$~~\\
\hline\hline
\multirow{2}{*}{$g\cdot1.5$}  & $E$    &  ~~$-19.25$~~  & ~~$15.41$~~ &~~$-18.71$~~ & ~~$0.04$~~& ~~$10630.75$~~\\
                                              & $E'$   &  ~~$-19.18$~~  & ~~$15.38$~~ &~~$-18.70$~~ & ~~-~~    & ~~$10630.82$~~\\
\hline\hline
\end{tabular}
\end{center}
\end{table}

\subsection{The $B^*\bar{B}^*$ system with $I^G=0^-$, $J^{PC}=1^{+-}$ }

We also investigate the $B^*\bar{B}^*$ system with $I^G=0^-$,
$J^{PC}=1^{+-}$. The effective potential and the meson-exchange
contribution are shown in Fig \ref{B-potential} and
\ref{B-potential-meson} A. The recoil correction is large. The
numerical results are listed in Table \ref{tab:BB0-}. The cutoff
varies from $1.0-1.3$ GeV. The binding energy with the recoil
correction is around $0-41.69$ MeV. The binding energy without the
recoil correction is between $0.3-53.47$ MeV. When the cutoff
parameter $\Lambda=1.0$ GeV, there is no bound state after adding
the recoil correction. The recoil correction is $0.3$ MeV. When the
cutoff parameter $\Lambda=1.3$ GeV, the binding energy is $41.69$
MeV. The recoil correction is $12.05$ MeV, and the contribution of
the spin-orbit force is $4.3$ MeV. In this case, the recoil
correction is significant and it is almost as big as the D-wave
contribution. But it is unfavorable for the formation of the
molecule.

\begin{table}[htbp]
\caption{The bound state solutions of the $B^* \bar{B}^{*}$ system
with $I^G=0^-$, $J^{PC}=1^{+-}$ (in unit of MeV) with the cutoff
$\Lambda$. The other notations are the same as in Table
\ref{tab:BB1+}.} \label{tab:BB0-}
\begin{center}
\begin{tabular}{c c c c c c | c}
\hline \hline \multirow{2}{*}{$\Lambda$(GeV)} & \multirow{2}{*}{~}  &\multicolumn{4}{c|}{Eigenvalue} & {Mass}  \\
\cline{2-6}
                                 &   &total   &S             &D        & LS   &(MeV) \\
\hline\hline
\multirow{2}{*}{1.0} & $E$    &  ~~-~~  & ~~-~~   & ~~-~~  & ~~~~     & ~~10650~~\\
                               & $E'$   &  ~~-0.30~~  & ~~-0.59~~   & ~~-0.07~~  & ~~-~~     & ~~10649.7~~\\
\hline\hline
\multirow{2}{*}{1.1} & $E$    &  ~~-4.34~~  & ~~-1.88~~   & ~~-3.08~~   & ~~0.42~~   & ~~10645.66~~\\
                               & $E'$   &  ~~-5.51~~  & ~~-2.19~~   & ~~-4.29~~   & ~~-~~   & ~~10644.49~~\\
\hline\hline
\multirow{2}{*}{1.2} & $E$    &  ~~-17.07~~  & ~~-9.05~~   & ~~-7.92~~     & ~~1.68~~   & ~~10632.93~~\\
                                 & $E'$   &  ~~-21.79~~  & ~~-12.09~~   & ~~-11.12~~    & ~~-~~     & ~~10628.21~~\\
\hline\hline
\multirow{2}{*}{1.3} & $E$    &  ~~-41.69~~  & ~~-24.61~~   & ~~-14.80~~     & ~~4.30~~      & ~~10608.31~~\\
                                 & $E'$  &  ~~-53.74~~  & ~~-30.81~~   & ~~-24.77~~     &  ~~-~~   & ~~10596.26~~\\
\hline\hline
\end{tabular}
\end{center}
\end{table}

\subsection{The $B^*\bar{B}^*$ system with $I^G=0^+$, $J^{PC}=1^{++}$ }

For the $B^*\bar{B}^*$ system with $I^G=0^+$, $J^{PC}=1^{++}$, $V_s$
and $V'_s$ are clearly different within the range of $0-0.6$ fm as
can be seen from Fig \ref{B-potential} B. Fig \ref{B-potential-meson}
B shows the contributions of each meson-exchange.
The recoil correction in
this system is important. The results are listed in Table
\ref{tab:BB0+}. When the variation of the cutoff parameter is
between $0.9-1.2$ GeV, the binding energy with the recoil correction
changes from $6.71$ to $59.74$ MeV. The binding energy without the
recoil correction changes from $6.86-63.76$ MeV. When the cutoff
parameter $\Lambda=0.9$ GeV, the binding energy is $6.71$ MeV. The
recoil correction is $0.15$ MeV, and the contribution of the
spin-orbit force is $0.03$ MeV. When the cutoff parameter
$\Lambda=1.2$ GeV, the binding energy is $63.76$ MeV. The recoil
correction is $4.02$ MeV, and the contribution of the spin-orbit
force is $1.39$ MeV. The recoil correction is of the same order as
the D-wave contribution. Moreover, it increase with the binding
energy. Therefore, the recoil effect can not be neglected. However,
the recoil correction is unfavorable for the formation of the
molecule in this system.

\begin{table}[htbp]
\caption{The bound state solutions of the $B^* \bar{B}^{*}$ system
with $I^G=0^+$, $J^{PC}=1^{++}$ (in unit of MeV) with the cutoff
$\Lambda$. The other notations are the same as in Table
\ref{tab:BB1+}.} \label{tab:BB0+}
\begin{center}
\begin{tabular}{c c c c c c | c}
\hline \hline \multirow{2}{*}{$\Lambda$(GeV)} & \multirow{2}{*}{~}  &\multicolumn{4}{c|}{Eigenvalue} & {Mass}  \\
\cline{2-6}
                                 &   &total   &S             &D        & LS   &(MeV) \\
\hline\hline
\multirow{2}{*}{0.9} & $E$    &  ~~-6.71~~  & ~~-22.27~~   & ~~0.18~~  & ~~0.03~~     & ~~10643.29~~\\
                               & $E'$   &  ~~-6.86~~  & ~~-22.85~~   & ~~0.15~~  & ~~-~~     & ~~10643.14~~\\
\hline\hline
\multirow{2}{*}{1.0} & $E$    &  ~~-19.36~~  & ~~-43.77~~   & ~~0.25~~  & ~~0.18~~     & ~~10630.64~~\\
                               & $E'$   &  ~~-20.15~~  & ~~-46.10~~   & ~~0.03~~  & ~~-~~     & ~~10629.85~~\\
\hline\hline
\multirow{2}{*}{1.1} & $E$    &  ~~-37.51~~  & ~~-65.91~~   & ~~0.53~~   & ~~0.58~~   & ~~10612.49~~\\
                               & $E'$   &  ~~-39.58~~  & ~~-71.13~~   & ~~-0.2~~   & ~~-~~   & ~~10610.42~~\\
\hline\hline
\multirow{2}{*}{1.2} & $E$    &  ~~-59.74~~  & ~~-86.12~~   & ~~1.34~~     & ~~1.39~~   & ~~10590.26~~\\
                                 & $E'$   &  ~~-63.76~~  & ~~-94.99~~   & ~~-0.43~~    & ~~-~~     & ~~10586.24~~\\
\hline\hline
\end{tabular}
\end{center}
\end{table}

\section{Numerical Results for $D^* \bar{D}^{*}$ system}\label{Numerical-DD}

\subsection{$D^* \bar{D}^{*}$ system with $I^G=1^+$, $J^{PC}=1^{+-}$}\label{DD1+}

Due to the isospin symmetry, the interaction in the $D^*
\bar{D}^{*}$ system has the same form with the $B^* \bar{B}^{*}$
system. Therefore, we repeat the same investigations for the $D^*
\bar{D}^{*}$ system. $Z_c(4025)$ was observed in the $\pi^{\mp}$
recoil mass spectrum in the process $e^-e^+ \rightarrow (D^*
\bar{D}^{*})^{\pm}\pi^{\mp}$ \cite{Ablikim:2014}. The mass of
$Z_c(4025)$ is close to the threshold of $D^* \bar{D}^{*}$, and
$Z_c(4025)$ has the quantum with $I^G=1^+$, $J^{PC}=1^{+-}$.
Therefore we first consider the possibility of the $D^* \bar{D}^{*}$
system as the molecular state with the quantum number $I^G=1^+$,
$J^{PC}=1^{+-}$.

From Fig \ref{D-potential} C, the curves of $V_s$ and $V'_s$ are
almost overlapping, same as those of $V_d$ and $V'_d$, which
indicates that the recoil correction is small. Fig
\ref{D-potential-meson} C shows that $\pi$ meson-exchange plays an
important role in the interaction.

Unfortunately, we did not get a bound state within a reasonable
range of the cutoff parameter and coupling constant. The value of
the pionic coupling constant was extracted from the decay width of
the $D^\ast$ meson where the pion is on the mass-shell. However we
need the value of the coupling constant in the potential where the
pion is off-shell. Considering the big influence of the $\pi$
meson-exchange, we multiply the coupling constant $g$ by a factor to
check the dependence of the results on $g$. The cutoff parameter is
fixed at $\Lambda =2.0$ GeV. The results are shown in Table
\ref{tab:DD1+}.

When the factor reaches $1.6$, there appears the bound state. The
binding energy with the recoil correction is $1.15$ MeV and the
recoil correction is $-0.05$ MeV. The contribution of the spin-orbit
force is $0.01$ MeV. When the factor is $1.8$, the binding energy
with recoil correction is $28.62$ MeV, and the recoil correction is
$-0.28$ MeV. The contribution of the spin-orbit force is $0.07$ MeV.
As in the $B^* \bar{B}^{*}$ system with $I^G=1^+$, $J^{PC}=1^{+-}$,
the recoil correction is not so big. But the recoil correction is
favorable for the formation of $D^* \bar{D}^{*}$ molecular state. In
other words, the existence of the $D^* \bar{D}^{*}$ molecule depends
on the coupling constant $g$ sensitively.

\begin{table}[htbp]
\caption{The $D^* \bar{D}^{*}$ system with $I^G=1^+$, $J^{PC}=1^{+-}$
(in units of MeV) with the variation of coupling constant $g$ and
$\Lambda =2.0$ GeV. The other notations are the same as in Table
\ref{tab:BB1+}.} \label{tab:DD1+}
\begin{center}
\begin{tabular}{c c c c c c | c}
\hline \hline \multirow{2}{*}{~$\Lambda $(GeV)~} & \multirow{2}{*}{~}  &\multicolumn{4}{c|}{Eigenvalue} & {Mass}   \\
\cline{2-6}
                                 &   &$Total$  &   $S$  &$D$   & $LS$ &(MeV) \\
\hline\hline
\multirow{2}{*}{$g\cdot1.6$}  & $E$    &  ~~$-1.15$~~  & ~~$-22.96$~~ & ~~$0.37$~~ & ~~$0.01$~~& ~~$4019.35$~~\\
                                               & $E'$   &  ~~$-1.10$~~  & ~~$-22.27$~~ &~~$0.36$~~ & ~~-~~& ~~$4019.40$~~\\
\hline\hline
\multirow{2}{*}{$g\cdot1.7$}   & $E$    &  ~~$-10.19$~~  & ~~$-75.21$~~ &~~$1.10$~~ & ~~$0.04$~~& ~~$4010.31$~~\\
                                               & $E'$   &  ~~$-10.03$~~  & ~~$-74.02$~~ &~~$1.08$~~ & ~~-~~   & ~~$4010.47$~~\\
\hline\hline
\multirow{2}{*}{$g\cdot1.8$}  & $E$    &  ~~$-28.62$~~  & ~~$-139.80$~~ &~~$1.80$~~ & ~~$0.07$~~& ~~$3991.88$~~\\
                                              & $E'$   &  ~~$-28.34$~~  & ~~$-138.16$~~ &~~$1.77$~~ & ~~-~~    & ~~$3992.16$~~\\
\hline\hline
\end{tabular}
\end{center}
\end{table}

\subsection{$D^* \bar{D}^{*}$ system with $I^G=1^-$, $J^{PC}=1^{++}$ }

We also investigate the $D^* \bar{D}^{*}$ system with $I^G=1^-$,
$J^{PC}=1^{++}$. From Fig \ref{D-potential} D and Fig
\ref{D-potential-meson} D, the recoil correction is very small while
the $\pi$ meson-exchange plays an significant role in the
interaction. There also does not exist a bound state when the cutoff
parameter is within a reasonable range. We also study the variation
with the coupling constant $g$. The results are shown in Table
\ref{tab:DD1-}.

The recoil correction is also very small in this system. For
example, when the binding energy is $4.71$ MeV, the recoil
correction is only $-0.03$ MeV, and the contribution of the
spin-orbit force is $0.05$ MeV. And the recoil correction is
favorable for the formation of the molecular state.

\begin{table}[htbp]
\caption{The $D^* \bar{D}^{*}$ system with $I^G=1^-$, $J^{PC}=1^{++}$
(in units of MeV) with the variation of coupling constant $g$ and
$\Lambda =2.0$ GeV. The other notations are the same as in Table
\ref{tab:BB1+}.} \label{tab:DD1-}
\begin{center}
\begin{tabular}{c c c c c c | c}
\hline \hline \multirow{2}{*}{~$\Lambda $(GeV)~} & \multirow{2}{*}{~}  &\multicolumn{4}{c|}{Eigenvalue} & {Mass}   \\
\cline{2-6}
                                 &   &$Total$  &   $S$  &$D$   & $LS$ &(MeV) \\
\hline\hline
\multirow{2}{*}{$g\cdot2.4$}  & $E$    &  ~~$-4.71$~~  & ~~$7.72$~~ & ~~$-8.47$~~ & ~~$0.05$~~& ~~$4015.79$~~\\
                                               & $E'$   &  ~~$-4.68$~~  & ~~$7.69$~~ &~~$-8.49$~~ & ~~-~~& ~~$4015.82$~~\\
\hline\hline
\multirow{2}{*}{$g\cdot2.5$}   & $E$    &  ~~$-11.05$~~  & ~~$13.22$~~ &~~$-14.96$~~ & ~~$0.09$~~& ~~$4009.45$~~\\
                                               & $E'$   &  ~~$-11.02$~~  & ~~$13.19$~~ &~~$-15.02$~~ & ~~-~~   & ~~$4009.48$~~\\
\hline\hline
\multirow{2}{*}{$g\cdot2.6$}  & $E$    &  ~~$-20.37$~~  & ~~$19.46$~~ &~~$?22.76$~~ & ~~$0.13$~~& ~~$4000.13$~~\\
                                              & $E'$   &  ~~$-20.34$~~  & ~~$19.43$~~ &~~$-22.86$~~ & ~~-~~    & ~~$4000.16$~~\\
\hline\hline
\end{tabular}
\end{center}
\end{table}

\subsection{The $D^* \bar{D}^{*}$ system with $I^G=0^-$, $J^{PC}=1^{+-}$ and $I^G=0^+$, $J^{PC}=1^{++}$}

For the isoscalar $D^* \bar{D}^{*}$ system, $V_s$ and $V'_s$ are
very different from Fig. \ref{D-potential} A and B. Fig \ref{D-potential-meson}
A and B show the contributions of each meson-exchange.
The recoil contribution is large and unfavorable for the formation of the
molecular states. There exist bound states for the isoscalar $D^*
\bar{D}^{*}$ system with different C-parity. The results are listed
in Tables \ref{tab:DD0-} and \ref{tab:DD0+}.

For the $J^{PC}=1^{+-}$ case, when cutoff parameter changes from
$1.4-1.6$ GeV, the binding energy with the recoil correction is
within $0.58-17.09$ MeV. The binding energy without the recoil
correction is within $3.83-17.09$ MeV. The recoil correction is
large. For example, when the binding energy is $0.58$ MeV, the
recoil correction is $3.25$ MeV, even bigger than the binding energy
itself. The contribution of the spin-orbit force is $0.4$ MeV, which
is almost as large as the binding energy.

For the $J^{PC}=1^{++}$ case, when cutoff parameter changes from
$1.3-1.6$ GeV, the binding energy with the recoil correction is
within $9.13-43.25$ MeV. The binding energy without the recoil
correction is within $10.59-49.23$ MeV. The recoil correction is
also significant. For example, when the binding energy is $9.13$
MeV, the recoil correction is $1.46$ MeV. The contribution of the
spin-orbit force is $0.75$ MeV, which is almost as large as D-wave
contribution. When the binding energy is $43.25$ MeV, the recoil
correction is $5.98$ MeV. The contribution of the spin-orbit force
is $5.21$ MeV.

\begin{table}[htbp]
\caption{The $D^* \bar{D}^{*}$ system with $I^G=0^-$, $J^{PC}=1^{+-}$
(in unit of MeV). The other notations are the same as in Table
\ref{tab:BB1+}.} \label{tab:DD0-}
\begin{center}
\begin{tabular}{c c c c c c | c}
\hline \hline \multirow{2}{*}{~$\Lambda (GeV)$~} & \multirow{2}{*}{~}  &\multicolumn{4}{c|}{Eigenvalue} & {Mass}   \\
\cline{2-6}
                                 &   &total  &   S  &D   & LS &($MeV$) \\
\hline\hline
\multirow{2}{*}{1.4}  & $E$    &  ~~-0.58~~  & ~~-5.03~~ & ~~-0.43~~ & ~~0.40~~& ~~4019.92~~\\
                                 & $E'$   &  ~~-3.83~~  & ~~-15.14~~ &~~-3.06~~ & ~~-~~& ~~4016.67~~\\
\hline\hline
\multirow{2}{*}{1.5}  & $E$    &  ~~-5.67~~  & ~~-19.23~~ &~~-1.45~~ & ~~1.91~~& ~~4014.83~~\\
                                & $E'$   &  ~~-17.25~~  & ~~-44.41~~ &~~-8.84~~ & ~~-~~& ~~4003.25~~\\
\hline\hline
\multirow{2}{*}{1.6}  & $E$    &  ~~-17.09~~  & ~~-40.12~~ &~~-2.52~~ & ~~4.76~~& ~~4003.41~~\\
                                & $E'$    & ~~-42.61~~   & ~~-79.61~~ &~~-17.73~~ & ~~-~~& ~~3977.89~~\\
\hline\hline
\end{tabular}
\end{center}
\end{table}

\begin{table}[htbp]
\caption{The $D^* \bar{D}^{*}$ system with $I^G=0^+$, $J^{PC}=1^{++}$
(in unit of MeV). The other notations are the same as in Table
\ref{tab:BB1+}.} \label{tab:DD0+}
\begin{center}
\begin{tabular}{c c c c c c | c}
\hline \hline \multirow{2}{*}{~$\Lambda (GeV)$~} & \multirow{2}{*}{~}  &\multicolumn{4}{c|}{Eigenvalue} & {Mass}   \\
\cline{2-6}
                                 &   &total  &   S  &D   & LS &($MeV$) \\
\hline\hline
\multirow{2}{*}{1.3}  & $E$    &  ~~-9.13~~  & ~~-34.43~~ & ~~1.04~~ & ~~0.75~~& ~~4011.37~~\\
                                 & $E'$   &  ~~-10.59~~  & ~~-42.22~~ &~~-0.15~~ & ~~-~~   & ~~4009.91~~\\
\hline\hline
\multirow{2}{*}{1.4}  & $E$    &  ~~-18.20~~  & ~~-49.42~~ & ~~2.25~~ & ~~1.66~~& ~~4002.3~~\\
                                 & $E'$   &  ~~-20.87~~  & ~~-61.74~~ &~~-0.29~~ & ~~-~~     & ~~3999.63~~\\
\hline\hline
\multirow{2}{*}{1.5}  & $E$    &  ~~-29.64~~  & ~~-63.30~~ &~~4.22~~ & ~~3.11~~& ~~3990.86~~\\
                                & $E'$   &  ~~-33.82~~  & ~~-80.59~~ &~~0.57~~ & ~~-~~      & ~~3986.68~~\\
\hline\hline
\multirow{2}{*}{1.6}  & $E$    &  ~~-43.25~~  & ~~-76.05~~ &~~7.17~~ & ~~5.21~~& ~~3977.25~~\\
                                & $E'$    & ~~-49.23~~   & ~~-98.50~~ &~~1.10~~ & ~~-~~     & ~~3971.27~~\\
\hline\hline
\end{tabular}
\end{center}
\end{table}

\section{Summary and Discussion}\label{summary}

With the one-boson-exchange model, we have systematically studied
the possible loosely bound $B^* \bar{B}^{*}$ and $D^* \bar{D}^{*}$
systems with (1) $I^G=0^+$, $J^{PC}=1^{++}$, (2) $I^G=0^-$,
$J^{PC}=1^{+-}$, (3) $I^G=1^+$, $J^{PC}=1^{+-}$ and (4) $I^G=1^-$,
$J^{PC}=1^{++}$. We consider the $\pi$, $\eta$ $\sigma$, $\rho$ and
$\omega$ meson exchange in the derivation of the potential. We keep
the momentum dependent terms in the polarization vector of the two
heavy mesons and introduce the momentum-related terms in the
interaction, which lead to the recoil correction and spin-orbit
force at $O(1/M^2)$.

The $B^* \bar{B}^{*}$ system with $I^G=1^+$, $J^{PC}=1^{+-}$ can
form the molecular state, which may correspond to the heavier $Z_b$
state observed by Belle collaboration. When the cutoff parameter is
within $2.2-2.6$ GeV, the binding energy is between $0.97-15.15$
MeV. The recoil correction is small. The contribution of the
spin-orbit force is also very small. For example, when the binding
energy is $15.15$ MeV, the recoil correction is $-0.17$ MeV. The
contribution of the spin-orbit force is $0.02$ MeV. But the recoil
correction is favorable to the formation of the molecular state. On
the other hand, our results shows that the binding energy is
sensitive to the pionic coupling constant.

For the isoscalar $B^* \bar{B}^{*}$ system, there exist a bound
state when changing the cutoff parameter. For the $J^{PC}=1^{+-}$
state, when cutoff parameter is within $1.0-1.3$ GeV, the binding
energy is between $0-41.69$ MeV. For the $J^{PC}=1^{++}$ state, when
cutoff parameter is within $0.9-1.2$ GeV, the binding energy is
between $6.71-59.74$ MeV. The recoil correction of the two systems
are both large and important. However, they are unfavorable to the
formation of the molecular states.

For the $D^* \bar{D}^{*}$ system with $I^G=1^+$, $J^{PC}=1^{+-}$, we
are unable to obtain the bound state within a reasonable cutoff
range and the pionic coupling constant $g$ extracted from the
$D^\ast $ decay width. If we enlarge the pionic coupling constant by
a factor of $1.6-1.8$, there appears the bound state with the
binding energy around $1.15-28.62$ MeV. The recoil correction is
small. For example, when the binding energy is $28.62$ MeV, the
recoil correction is $-0.28$ MeV. The contribution of the spin-orbit
force is $0.07$ MeV. The recoil correction is favorable for the
formation of the molecular state.

For the isoscalar $D^* \bar{D}^{*}$ system, there exist bound states
when changing the cutoff parameter. For the $J^{PC}=1^{+-}$ state,
when the cutoff parameter is within $1.4-1.6$ GeV, the binding
energy is around $0.58-42.61$ MeV. For the $J^{PC}=1^{++}$ state,
when the cutoff parameter is within $1.3-1.6$ GeV, the binding
energy is around $9.13-43.25$ MeV. The recoil correction is
significant but unfavorable to the formation of the molecular
states.

\section{Appendix}

We collect the lengthy formulae in the appendix.
\begin{equation}
Y(\tilde{m}_{\alpha}r)=\frac{\exp(\tilde{m}_{\alpha}r)}{\tilde{m}_{\alpha}r}
\end{equation}

\begin{equation}
Z(\tilde{m}_{\alpha}r)=(1+\frac{3}{\tilde{m}_{\alpha}r}+\frac{3}{(\tilde{m}_{\alpha}r)^2})Y(\tilde{m}_{\alpha}r)
\end{equation}

\begin{equation}
Z_1(\tilde{m}_{\alpha}r)=(\frac{1}{\tilde{m}_{\alpha}r}+\frac{1}{(\tilde{m}_{\alpha}r)^2})Y(\tilde{m}_{\alpha}r)
\end{equation}

\begin{equation}
Z'(\tilde{m}_{\alpha}r)=\frac{\sin(\tilde{m}_{\alpha}r)}{\tilde{m}_{\alpha}r}-\frac{3}{\tilde{m}_{\alpha}r}\frac{\sin(\tilde{m}_{\alpha}r)}{\tilde{m}_{\alpha}r}+\frac{1}{(\tilde{m}_{\alpha}r)^2}\frac{\cos(\tilde{m}_{\alpha}r)}{\tilde{m}_{\alpha}r}.
\end{equation}

\begin{equation}
Z'_1(\tilde{m}_{\alpha}r)=\frac{1}{\tilde{m}_{\alpha}r}\frac{\sin(\tilde{m}_{\alpha}r)}{\tilde{m}_{\alpha}r}+\frac{1}{(\tilde{m}_{\alpha}r)^2}\frac{\cos(\tilde{m}_{\alpha}r)}{\tilde{m}_{\alpha}r}
\end{equation}
where for system $D \bar{D}^*$
\begin{equation}
\tilde{m}^2_{\pi}=(m_D^*-m_D)^2-m^2_{\pi},
\end{equation}
\begin{equation}
\tilde{m}^2_{\sigma,\rho,\omega,\eta}=m^2_{\sigma,\rho,\omega,\eta}-(m_D^*-m_D)^2.
\end{equation}
for system $B \bar{B}^*$
\begin{equation}
\tilde{m}^2_{\pi,\sigma,\rho,\omega,\eta}=m^2_{\pi,\sigma,\rho,\omega,\eta}-(m_B^*-m_B)^2.
\end{equation}

\begin{eqnarray}
\mathcal{F}_{1t\alpha}&=&\mathcal{F}\{(\frac{\Lambda^2-m_{\alpha}^2}{{\Lambda}^2+\vec{q}^2})
\frac{1}{\vec{q}^2+m_{\alpha}^2}\}\nonumber\\
&=&m_{\alpha}Y(m_{\alpha}r)-\Lambda Y(\Lambda r)-
(\Lambda^2-m_{\alpha}^2)\frac{e^{-\Lambda r}}{2\Lambda}
\end{eqnarray}

\begin{eqnarray}
\mathcal{F}_{1u\alpha}&=&\mathcal{F}\{(\frac{\Lambda^2-m_{\alpha}^2}{\tilde{\Lambda}^2+\vec{q}^2})
\frac{1}{\vec{q}^2+\tilde{m}_{\alpha}^2}\}\nonumber\\
&=&\tilde{m}_{\alpha}Y(\tilde{m}_{\alpha}r)-\tilde{\Lambda}
Y(\tilde{\Lambda}
r)-(\Lambda^2-m_{\alpha}^2)\frac{e^{-\tilde{\Lambda}
r}}{2\tilde{\Lambda}}
\end{eqnarray}

\begin{eqnarray}
\mathcal{F}_{2t\alpha}&=&\mathcal{F}\{(\frac{\Lambda^2-m_{\alpha}^2}{{\Lambda}^2+\vec{q}^2})
\frac{\vec{q}^2}{\vec{q}^2+m_{\alpha}^2}\}\nonumber\\
&=&m_{\alpha}^2[\Lambda Y(\Lambda r)-m_{\alpha}Y(m_{\alpha}r)]\nonumber\\
&+&(\Lambda^2-m_{\alpha}^2)\Lambda\frac{e^{-\Lambda r}}{2}
\end{eqnarray}

\begin{eqnarray}
\mathcal{F}_{2u\alpha}&=&\mathcal{F}\{(\frac{\Lambda^2-m_{\alpha}^2}{{\tilde{\Lambda}}^2+\vec{q}^2})
\frac{\vec{q}^2}{\vec{q}^2+\tilde{m}_{\alpha}^2}\}\nonumber\\
&=&\tilde{m}_{\alpha}^2[\tilde{\Lambda} Y(\tilde{\Lambda} r)-\tilde{m}_{\alpha}Y(\tilde{m}_{\alpha}r)]\nonumber\\
&+&(\Lambda^2-m_{\alpha}^2)\tilde{\Lambda}\frac{e^{-\tilde{\Lambda}
r}}{2}
\end{eqnarray}

\begin{eqnarray}
\mathcal{F}_{3t\alpha} &=&
\mathcal{F}\{(\frac{\Lambda^2-m_{\alpha}^2}{{\Lambda}^2+\vec{q}^2})
\frac{(\vec{\sigma_1}\cdot\vec{q})(\vec{\sigma_2}\cdot\vec{q})}{\vec{p}^2+m_{\alpha}^2}\}\nonumber\\
&=&\frac{1}{3}\vec{\sigma_1}\cdot\vec{\sigma_2}[~m_{\alpha}^2\Lambda Y(\Lambda r)-m_{\alpha}^3 Y(m_{\alpha}r)\nonumber\\
&+&(\Lambda^2-m_{\alpha}^2)\Lambda\frac{e^{-\Lambda r}}{2}~]\nonumber\\
&+&\frac{1}{3}S_{12}[-m_{\alpha}^3 Z(m_{\alpha}r)+ \Lambda^3 Z(\Lambda r) \nonumber\\
&+&(\Lambda^2-m_{\alpha}^2)(1+\Lambda r)
\frac{\Lambda}{2}Y(\Lambda r)]\nonumber\\
&=&(\vec{\sigma_1}\cdot\vec{\sigma_2})\mathcal{F}_{3t1} +
S_{12}\mathcal{F}_{3t2}
\end{eqnarray}

\begin{eqnarray}
\mathcal{F}_{3u\alpha} &=&
\mathcal{F}\{(\frac{\Lambda^2-m_{\alpha}^2}{{\tilde{\Lambda}}^2+\vec{q}^2})
\frac{(\vec{\sigma_1}\cdot\vec{q})(\vec{\sigma_2}\cdot\vec{q})}{\vec{q}^2+\tilde{m}_{\alpha}^2}\}\nonumber\\
&=&\frac{1}{3}\vec{\sigma_1}\cdot\vec{\sigma_2}[\tilde{m}_{\alpha}^2\tilde{\Lambda} Y(\tilde{\Lambda} r)-\tilde{m}_{\alpha}^3Y(\tilde{m}_{\alpha}r)\nonumber\\
&+&(\Lambda^2-m_{\alpha}^2)\tilde{\Lambda}\frac{e^{-\tilde{\Lambda} r}}{2}]\nonumber\\
&+&\frac{1}{3}S_{12}[-\tilde{m_{\alpha}}^3 Z(\tilde{m_{\alpha}}r)+ \tilde{\Lambda}^3 Z(\tilde{\Lambda} r) \nonumber\\
&+&(\Lambda^2-m_{\alpha}^2)(1+\tilde{\Lambda} r)\frac{\tilde{\Lambda}}{2}Y(\tilde{\Lambda} r)~]\nonumber\\
&=&(\vec{\sigma_1}\cdot\vec{\sigma_2})\mathcal{F}_{3u1\alpha} +
S_{12}\mathcal{F}_{3u2\alpha}
\end{eqnarray}

\begin{eqnarray}
\mathcal{F'}_{3u\alpha}&=&
\mathcal{F}\{(\frac{\Lambda^2-m_{\alpha}^2}{{\tilde{\Lambda}}^2+\vec{q}^2})
\frac{(\vec{\sigma_1}\cdot\vec{q})(\vec{\sigma_2}\cdot\vec{q})}{\vec{p}^2-\tilde{m_{\alpha}}^2}\}\nonumber\\
&=&\frac{1}{3}\vec{\sigma_1}\cdot\vec{\sigma_2}[~-\tilde{m}_{\alpha}^2\tilde{\Lambda} Y(\tilde{\Lambda} r)-\tilde{m}_{\alpha}^3\frac{\cos(\tilde{m}_{\alpha}r)}{\tilde{m}_{\alpha}r}\nonumber\\
&+&(\Lambda^2-m_{\alpha}^2)\tilde{\Lambda}\frac{e^{-\tilde{\Lambda} r}}{2}~]\nonumber\\
&+&\frac{1}{3}S_{12}[\tilde{m}_{\alpha}^3 Z'(\tilde{m}_{\alpha}r)+ \tilde{\Lambda}^3 Z(\tilde{\Lambda} r)\nonumber\\
&+&(\Lambda^2-m_{\alpha}^2)(1+\tilde{\Lambda} r)\frac{\tilde{\Lambda}}{2}Y(\tilde{\Lambda} r)~]\nonumber\\
&=&(\vec{\sigma_1}\cdot\vec{\sigma_2})\mathcal{F'}_{3u1\alpha} +
S_{12}\mathcal{F'}_{3u2\alpha}
\end{eqnarray}

\begin{eqnarray}
\mathcal{F}_{4t\alpha}
&=&\mathcal{F}\{{(\frac{\Lambda^2-m_{\alpha}^2}{{\Lambda}^2+\vec{q}^2})
\frac{\vec{k}^2}{\vec{q}^2+m_{\alpha}^2}}\}\nonumber\\
&=&\frac{m_{\alpha}^3}{4}Y(m_{\alpha}r)-\frac{\Lambda^3}{4}Y(\Lambda r)\nonumber\\
&-&\frac{\Lambda^2-m_{\alpha}^2}{4}(\frac{\Lambda r}{2}-1)\frac{e^{-\Lambda r}}{r}\nonumber\\
&-&\frac{1}{2}\{\nabla^2,m_{\alpha} Y(m_{\alpha}r)-\Lambda Y(\Lambda r)-\frac{\Lambda^2-m_{\alpha}^2}{2}\frac{e^{-\Lambda r}}{\Lambda}\}\nonumber\\
&=&\mathcal{F}_{4t1\alpha}+\{-\frac{1}{2}\nabla^2,\mathcal{F}_{4t2\alpha}\}
\end{eqnarray}

\begin{eqnarray}
\mathcal{F}_{4u\alpha}
&=&\mathcal{F}\{{(\frac{\Lambda^2-\tilde{m}_{\alpha}^2}{{\tilde{\Lambda}}^2+\vec{q}^2})
\frac{\vec{k}^2}{\vec{q}^2+\tilde{m}_{\alpha}^2}}\}\nonumber\\
&=&\frac{\tilde{m}_{\alpha}^3}{4}Y(\tilde{m}_{\alpha}r)-\frac{\tilde{\Lambda}^3}{4}Y(\tilde{\Lambda} r)\nonumber\\
&-&\frac{\Lambda^2-m_{\alpha}^2}{4}(\frac{\tilde{\Lambda} r}{2}-1)\frac{e^{-\tilde{\Lambda} r}}{r}\nonumber\\
&-&\frac{1}{2}\{\nabla^2,\tilde{m}_{\alpha}Y(\tilde{m}_{\alpha}r)-\tilde{\Lambda}
Y(\tilde{\Lambda} r)-\frac{\Lambda^2-m_{\alpha}^2}{2}
\frac{e^{-\tilde{\Lambda} r}}{\tilde{\Lambda}}\}\nonumber\\
&=&\mathcal{F}_{4u1\alpha}+\{-\frac{1}{2}\nabla^2,\mathcal{F}_{4u2\alpha}\}
\end{eqnarray}

\begin{eqnarray}
\mathcal{F}_{5t\alpha}
&=&\mathcal{F}\{{i(\frac{\Lambda^2-m_{\alpha}^2}{{\Lambda}^2+\vec{q}^2})
\frac{\vec{S}\cdot(\vec{q}\times\vec{k})}{\vec{q}^2+m_{\alpha}^2}}\}\nonumber\\
&=&\vec{S}\cdot\vec{L}[-m_{\alpha}^3 Z_{1}(m_{\alpha}r)+\Lambda^3 Z_{1}(\Lambda r)\nonumber\\
&+&(\Lambda^2-m_{\alpha}^2)\frac{e^{-\Lambda r}}{2r}]\nonumber\\
&=&\vec{S}\cdot\vec{L}\mathcal{F}_{5t0\alpha}
\end{eqnarray}

\begin{eqnarray}
\mathcal{F}_{5u\alpha}
&=&\mathcal{F}\{{i(\frac{\Lambda^2-m_{\alpha}^2}{{\tilde{\Lambda}}^2+\vec{q}^2})
\frac{\vec{S}\cdot(\vec{q}\times\vec{k})}{\vec{q}^2+\tilde{m}_{\alpha}^2}}\}\nonumber\\
&=&\vec{S}\cdot\vec{L}[-\tilde{m}_{\alpha}^3 Z_{1}(\tilde{m}_{\alpha}r)+\tilde{\Lambda}^3 Z_{1}(\tilde{\Lambda} r)\nonumber\\
&+&(\Lambda^2-m_{\alpha}^2)\frac{e^{-\tilde{\Lambda} r}}{2r}]\nonumber\\
&=&\vec{S}\cdot\vec{L}\mathcal{F}_{5u0\alpha}
\end{eqnarray}

\begin{eqnarray}
\mathcal{F'}_{5u\alpha}
&=&\mathcal{F}\{{i(\frac{\Lambda^2-m_{\alpha}^2}{{\tilde{\Lambda}}^2+\vec{q}^2})
\frac{\vec{S}\cdot(\vec{q}\times\vec{k})}{\vec{q}^2+\tilde{m_{\alpha}}^2}}\}\nonumber\\
&=&\vec{S}\cdot\vec{L}[-\tilde{m}_{\alpha}^3 Z'_{1}(\tilde{m}_{\alpha}r)+\tilde{\Lambda}^3 Z_{1}(\tilde{\Lambda} r)\nonumber\\
&+&(\Lambda^2-m_{\alpha}^2)\frac{e^{-\tilde{\Lambda} r}}{2r}]\nonumber\\
&=&\vec{S}\cdot\vec{L}\mathcal{F'}_{5u0\alpha}
\end{eqnarray}

\begin{eqnarray}
\mathcal{F}_{6u\alpha}&=&
\mathcal{F}\{(\frac{\Lambda^2-m_{\alpha}^2}{{\tilde{\Lambda}}^2+\vec{q}^2})
\frac{(\vec{\sigma_1}\cdot\vec{k})(\vec{\sigma_2}\cdot\vec{k})}{\vec{p}^2+\tilde{m}_{\alpha}^2}\}\nonumber\\
&=&-\frac{\vec{\sigma_1}\cdot\vec{\sigma_2}}{4}[~\tilde{m}_{\alpha}^3Y(\tilde{m}_{\alpha}r)-(\tilde{\Lambda})^3Y(\tilde{\Lambda}r)\nonumber\\
&-&(\Lambda^2-m_{\alpha}^2)\tilde{\Lambda}\frac{e^{-\tilde{\Lambda} r}}{2}~]\nonumber\\
&+&\frac{1}{3}(S_{12}+\vec{\sigma_1}\cdot\vec{\sigma_2})[~(1+\frac{3}{\tilde{m}_{\alpha}r})\tilde{m}_{\alpha}^2 Y(\tilde{\Lambda} r)\nonumber\\
&-&(1+\frac{3}{\tilde{\Lambda}r})(\tilde{\Lambda})^2 Y(\tilde{\Lambda} r)\nonumber\\
&-&(\Lambda^2-m_{\alpha}^2)(\tilde{\Lambda}+\frac{2}{r})\frac{e^{-\tilde{\Lambda} r}}{2\tilde{\Lambda}}~]\nabla\nonumber\\
&-&\frac{1}{3}(S_{12}+\vec{\sigma_1}\cdot\vec{\sigma_2})[\tilde{m}_{\alpha} Y(\tilde{m}_{\alpha}r)- \tilde{\Lambda}Y(\tilde{\Lambda} r)\nonumber\\
&-&(\Lambda^2-m_{\alpha}^2)\frac{e^{-\tilde{\Lambda} r}}{2\tilde{\Lambda}}~]\nabla^2\nonumber\\
&=&-\frac{\vec{\sigma_1}\cdot\vec{\sigma_2}}{4}\mathcal{F}_{6u1\alpha}
+\frac{1}{3}(S_{12}+\vec{\sigma_1}\cdot\vec{\sigma_2})\mathcal{F}_{6u2\alpha}\nonumber\\
&-&\frac{1}{3}(S_{12}+\vec{\sigma_1}\cdot\vec{\sigma_2})\mathcal{F}_{6u3\alpha}
\end{eqnarray}

\begin{eqnarray}
\mathcal{F'}_{6u\alpha}&=&
\mathcal{F}\{(\frac{\Lambda^2-m_{\alpha}^2}{{\tilde{\Lambda}}^2+\vec{q}^2})
\frac{(\vec{\sigma_1}\cdot\vec{k})(\vec{\sigma_2}\cdot\vec{k})}{\vec{p}^2-\tilde{m}_{\alpha}^2}\}\nonumber\\
&=&-\frac{\vec{\sigma_1}\cdot\vec{\sigma_2}}{4}[~\tilde{m}_{\alpha}^3 \frac{\cos(M_{\alpha}r)}{\tilde{m}_{\alpha}r}-(\tilde{\Lambda})^3Y(\tilde{\Lambda}r)\nonumber\\
&-&(\Lambda^2-m_{\alpha}^2)\tilde{\Lambda}\frac{e^{-\tilde{\Lambda} r}}{2}~]\nonumber\\
&+&\frac{1}{3}(S_{12}+\vec{\sigma_1}\cdot\vec{\sigma_2})[~(\frac{\sin(\tilde{m}_{\alpha}r)}{\tilde{m}_{\alpha}r}+\frac{3}{\tilde{m}_{\alpha}r}\frac{\cos(\tilde{m}_{\alpha}r)}
{\tilde{m}_{\alpha}r})\tilde{m}_{\alpha}^2\nonumber\\
&-&(1+\frac{3}{\tilde{\Lambda}r})(\tilde{\Lambda})^2
Y(\tilde{\Lambda} r)
-(\Lambda^2-m_{\alpha}^2)(\tilde{\Lambda}+\frac{2}{r})\frac{e^{-\tilde{\Lambda} r}}{2\tilde{\Lambda}}~]\nabla\nonumber\\
&-&\frac{1}{3}(S_{12}+\vec{\sigma_1}\cdot\vec{\sigma_2})[\tilde{m}_{\alpha} \frac{\cos(\tilde{m}_{\alpha}r)}{\tilde{m}_{\alpha}r}- \tilde{\Lambda}Y(\tilde{\Lambda} r)\nonumber\\
&-&(\Lambda^2-m_{\alpha}^2)\frac{e^{-\tilde{\Lambda} r}}{2\tilde{\Lambda}}~]\nabla^2\nonumber\\
&=&-\frac{\vec{\sigma_1}\cdot\vec{\sigma_2}}{4}\mathcal{F'}_{6u1\alpha}
+\frac{1}{3}(S_{12}+\vec{\sigma_1}\cdot\vec{\sigma_2})\mathcal{F'}_{6u2\alpha}\nonumber\\
&-&\frac{1}{3}(S_{12}+\vec{\sigma_1}\cdot\vec{\sigma_2})\mathcal{F'}_{6u3\alpha}
\end{eqnarray}

\section*{ACKNOWLEDGEMENT}

We thank Li-Ping Sun for useful discussions. This project is
supported by the National Natural Science Foundation of China under
Grant No. 11261130311.


\begin{thebibliography}{50}

\bibitem{Choi:2003}
S.K. Choi {\it et~al.}, Belle Collaboration,
 \newblock Phys. Rev. Lett. {\bf 91}, 262001 (2003).

\bibitem{B. Aubert:2005}
B. Aubert {\it et~al.}, $BARBAR$ Collaboration,
\newblock Phys. Rev. Lett. {\bf 95}, 142001 (2005).

\bibitem{C.Z. Yuan:2007}
C.Z. Yuan {\it et~al.}, Belle Collaboration,
\newblock Phys. Rev. Lett. {\bf 99}, 182001 (2007).

\bibitem{B. Aubert:2007}
B. Aubert {\it et~al.}, $BARBAR$ Collaboration,
\newblock Phys. Rev. Lett. {\bf 98}, 212001 (2007).

\bibitem{X.L. Wang:2007}
X.L. Wang {\it et~al.}, Belle Collaboration,
\newblock Phys. Rev. Lett. {\bf 99}, 142002 (2007).

\bibitem{G.Pakhlova:2008}
G.Pakhlova {\it et~al.}, Belle Collaboration,
\newblock Phys. Rev. Lett. {\bf 101}, 172001 (2008).

\bibitem{R.Mizuk:2008}
R.~Mizuk {\it et~al.}, Belle Collaboration,
\newblock Phys. Rev. {\bf D78}, 072004 (2008).

\bibitem{Choi:2008}
S.K. Choi  {\it et~al.}, Belle Collaboration,
\newblock Phys. Rev. Lett. {\bf 100}, 142001 (2008).

\bibitem{K.Chilikin:2013}
K.Chilikin  {\it et~al.}, Belle Collaboration,
\newblock Phys. Rev. {\bf D88}, 074026 (2013).


\bibitem{Ablikim:2013}
M.~Ablikim  {\it et~al.}, BESIII Collaboration,
 \newblock Phys. Rev. Lett. {\bf 110}, 252001 (2013).

\bibitem{Z.Q. Liu:2013-2}
Z.Q. Liu {\it et~al.}, Belle Collaboration,
\newblock Phys. Rev. Lett. {\bf 110}, 252002 (2013).

\bibitem{T.Xiao:2013}
T. Xiao, S. Dobbs, A. Tomaradze and Kamal K. Seth,
 \newblock Phys. Lett. {\bf B727}, 366 (2013).

\bibitem{M. Ablikim:2013}
M.~Ablikim  {\it et~al.}, BESIII Collaboration,
 \newblock Phys. Rev. Lett. {\bf 111}, 242001 (2013).

\bibitem{Ablikim:2014}
M.~Ablikim  {\it et~al.}, BESIII Collaboration,
 \newblock Phys. Rev. Lett. {\bf 112}, 132001 (2014).

\bibitem{I.Adachi:2011}
I.~Adachi  {\it et~al.}, Belle Collaboration,
 \newblock arXiv:1105.4583 [hep-ex]


\bibitem{ZHUPLB2005}
S. L. Zhu, Phys.Lett. B. {\bf 625}, 212 (2005).

\bibitem{H.Hogaasen:2006}
H. Hogaasen, J.M. Richard and P. Sorba,
\newblock Phys. Rev. {\bf D73}, 054013 (2006).

\bibitem{D.Ebert:2006}
D. Ebert, R.N. Faustov and V.O. Galkin,
\newblock Phys. Lett. {\bf B634}, 214 (2006).

\bibitem{N.Barnea:2006}
N. Barnea, J. Vijande and A. Valcarce,
\newblock Phys. Rev. {\bf D73}, 054004 (2006).

\bibitem{Y.Cui:2007}
Y. Cui, X.L. Chen, W.Z. Deng and S.L. Zhu,
 \newblock High Energy Phys. Nucl. Phys. {\bf 31}, 7 (2007).

\bibitem{R.D.Matheus:2007}
R.D. Matheus, S. Narison, M. Nielsen and J.M. Richard,
 \newblock Phys. Rev. {\bf D75}, 014005 (2007).

\bibitem{T.W.Chiu:2007}
T.W.Chiu and T.H. Hsieh,
\newblock Phys. Lett. {\bf B646}, 95 (2007).

\bibitem{L.Zhao:2014}
L. Zhao, W.Z. Deng and S.L. Zhu,
\newblock Phys.Rev. {\bf D90}, 094031 (2014).


\bibitem{D.Gamermann:2007}
D.~Gamermann and E.~Oset,
 \newblock Eur. Phys. J. A {\bf 33}, 119 (2007).

\bibitem{F.E.Close:2004}
F.~E.~Close and P.R.~Page,
\newblock Phys. Lett. B {\bf 578}, 119 (2004).

\bibitem{M.B.Voloshin:2004}
M.~B.~Voloshin,
\newblock Phys. Lett. B {\bf 579}, 316 (2004).

\bibitem{C.Y.Wong:2004}
C.~Y.~Wong,
\newblock Phys. Rev. C {\bf 69}, 055202 (2004).

\bibitem{E.S.Swanson:2004}
E.~S.~Swanson,
\newblock Phys. Lett. B {\bf 588}, 189 (2004).

\bibitem{N.A.Tornqvist:2004}
N.~A.~T\"{o}rnqvist,
\newblock Phys. Lett. B {\bf 590}, 209 (2004).

\bibitem{Y.R.Liu:2010}
Y.-R.~Liu, M. Oka, M.~Takizawa, X.~Liu, W.-Z.~Deng, and S.-L.~Zhu,
\newblock Phys. Rev. D {\bf 82}, 014011 (2010).

\bibitem{N.Li:2012}
Ning Li, Shi-Lin Zhu,
\newblock Phys.Rev. {\bf D86}, 074022 (2012).

\bibitem{mali}
L. Ma, X.-H. Liu, X. Liu, and S.-L. Zhu,
\newblock arXiv:1404.3450 [hep-ph].

\bibitem{liuxiaohai}
X.-H. Liu, L. Ma, L.-P. Sun, X. L., and S.-L. Zhu,
\newblock arXiv:1407.3684 [hep-ph].


\bibitem{Q.Wang:2013}
Q. Wang, C. Hanhart and Q. Zhao,
\newblock arXiv:1303.6355 [hep-ph]

\bibitem{Oset}
F. Aceti, M. Bayar, E. Oset, A. Martinez Torres, K. P. Khemchandani,
F. S. Navarra and M. Nielsen,
\newblock arXiv:1401.8216 [hep-ph]

\bibitem{Y.R.Liu:2008}
Y.-R.Liu, X.Liu, W.-Z. Deng and S.-L. Zhu,
 \newblock Eur. Phys. J. {\bf C56}, 63 (2008).

\bibitem{X.Liu:2009}
X.Liu, L.-Z. Gang, Y.-R.Liu, and S.-L. Zhu,
 \newblock Eur. Phys. J. {\bf C61}, 411 (2009).

\bibitem{A.E.Bondar:2011}
A.E. Bondar, A. Garmash, A.I. Milstein, R. Mizuk and M.B. Voloshin,
\newblock arXiv:1105.4437 [hep-ph]

\bibitem{L.Xiang:2011}
Z.-F. Sun, J. He, X. Liu, Z.-G. Luo, and S.-L. Zhu,
\newblock Phys.Rev. {\bf D84}, 054002 (2011).


\bibitem{W.Chen:2014}
W. Chen, T.G. Steele, M.-L. Du, and S.-L. Zhu,
\newblock Eur.Phys.J. {\bf C74}, 2773 (2014).

\bibitem{J.He:2013}
J. He, X. Liu, Z.-F. Sun, and S.-L. Zhu,
\newblock Eur.Phys.J. {\bf C73}, 2635 (2013).

\bibitem{Z.F.Sun:2012}
Z.-F. Sun, Z.-G. Luo, J. He, X. Liu, and S.-L. Zhu,
\newblock Chin.Phys. {\bf C36}, 194 (2012).

\bibitem{C.D. Deng:2014}
C.D. Deng:2014, J.L. Ping and F. Wang,
\newblock arXiv:1402.0777 [hep-ph]

\bibitem{J.M. Dias:2014}
J. M. Dias, F. S. Navarra, M. Nielsen and C. Zanetti,
\newblock arXiv:1311.7591s [hep-ph]


\bibitem{L.Zhao:2013}
L. Zhao, L. Ma, and S.-L. Zhu,
\newblock Phys.Rev. {\bf D89}, 094026 (2014).

\bibitem{S.Ahmed:2001}
S.~Ahmed  {\it et~al.}, (CLEO Collaboration),
\newblock Phys. Rev. Lett. {\bf 87}, 251801 (2001).

\bibitem{C.Isola:2003}
C.~Isola, M.~Ladisa, G.~Nardulli, and P.~Santorelli,
\newblock Phys. Rev. D {\bf 68}, 114001 (2003).

\bibitem{M.Bando:1988}
C.~Isola, M.~Ladisa, G.~Nardulli, and P.~Santorelli,
\newblock Phys. Rep. {\bf 164}, 217 (1988).

\bibitem{A.F.Falk:1992}
A.~F.~Falk and M.~E.~Luke,
\newblock Phys. Lett. B {\bf 292}, 119 (1992).

\bibitem{PDG}
K.~Nakamura, {\it et~al.}, (Particle Data Group),
\newblock J. Phys. G {\bf 37}, 075021 (2010).


\end{thebibliography}
\end{document}